# Analysis and Design of Desaturation Short-Circuit Protection for SiC MOSFET Under High Switching Voltage Transients

Ravi Teja Pogulaguntla, *Graduate Student Member, IEEE* and Arun Karuppaswamy B, *Senior Member, IEEE*

***Abstract*—Short-circuit protection is vital for ensuring the safe operation of high-power-density power converters using Silicon Carbide (SiC) MOSFETs. However, adapting conventional desaturation (DESAT) protection, originally designed for Silicon (Si) IGBTs, is challenging due to the significantly shorter short-circuit withstand time of SiC devices, typically less than 3 $\mu$s. Nevertheless, DESAT protection can be adapted for SiC MOSFETs through certain circuit modifications. This article provides a comprehensive analysis and design guidelines for the three architectural variants of DESAT protection circuits, specifically for adjusting the short-circuit detection time suitable for SiC MOSFETs. Further, this article establishes a quantitative framework to evaluate the noise induced into the DESAT protection circuits during turn-off $dv/dt$ switching transients. From this analysis, a design methodology and safe operating boundary are derived for each DESAT configuration, guiding the systematic selection of component values to balance $dv/dt$ noise margin and short-circuit detection speed for SiC MOSFETs. Experimental results are presented to validate the analysis.**

## I. Introduction

Silicon Carbide (SiC) MOSFETs offer significantly faster switching speeds compared to Silicon (Si) IGBTs. Despite this advantage, their output characteristics pose severe challenges during a short-circuit fault. Specifically, the peak short-circuit current in a SiC MOSFET can reach up to ten times its nominal current rating [1], because of the lack of high saturation of current in the active region. The ohmic effect in an SiC MOSFET extends well beyond the typical $6 - 9$ V window observed in Si-IGBTs, with a different slope. This current surge induces a much higher instantaneous power dissipation within a semiconductor chip area that is significantly smaller than that of a Si counterpart. The higher power dissipation offsets the better thermal conductivity of SiC and requires a much shorter short-circuit withstand time for SiC MOSFETs than Si IGBTs [2]. Therefore, it is essential to implement a faster short-circuit protection for SiC MOSFETs to ensure the safe operation of SiC-based power converters.

Several short-circuit protection methods have been reported in the literature for SiC MOSFETs. These methodologies can be broadly categorized into three fundamental groups: current-sensing methods [3], [4], [5], [6], [7], gate-charge characteristic tracking method [8], and voltage-sensing methods [9], [10]. Specifically, to detect the short-circuit fault, current-sensing-based methods rely on shunt resistors or current transformers to directly monitor the device current. The gate-charge tracking methods monitor the gate-voltage profile and gate charge. The voltage-sensing-based techniques infer a short-circuit condition by directly or indirectly monitoring the voltage across the device. A discussion of the prior literature on the three approaches is presented as follows.

A current transformer followed by signal conditioning circuitry is utilized in literature [3] and [4] to detect the short-circuit fault. However, current transformers increase overall volume and are better suited for integration into SiC power modules than discrete devices. In [5], a Sense MOSFET and a Main MOSFET are integrated into a single package. The Sense MOSFET acts as a current mirror for the Main MOSFET but carries a much lower current. The sense current is converted into an equivalent voltage using a shunt resistor and is subsequently utilized to detect short-circuit faults. However, this method poses challenges in device fabrication. Furthermore, larger shunt resistor values can alter the current-mirror ratio, thereby degrading detection accuracy. In [6], a pre-set peak fault current is estimated by passively integrating the change in drain-source current ($di_{ds}/dt$) induced voltage across the stray inductance ($L_{S_s}$) between the Kelvin-source and power source terminals to detect a short-circuit fault. The accuracy of this method is affected by the precise extraction of the $L_{S_s}$ value and by the design of the passive integrator for accurately measuring high $di_{ds}/dt$, which is typical for SiC MOSFETs. In [7], a high-bandwidth Rogowski coil embedded within a printed circuit board (PCB) is designed to detect short-circuit faults. However, this method requires additional PCB footprint to accommodate the coil geometry, thereby increasing the power loop stray inductance ($L_S$). The increased $L_S$ aggravates the voltage overshoot across the device during turn-off and prolongs switching transients. Consequently, this approach increases the overall switching energy losses of the device [11].

In [8], the short-circuit fault is detected by monitoring the gate voltage profile and measuring the gate charge. In this method, establishing accurate reference thresholds for both the gate voltage and gate charge is challenging in the fast-switching environment of SiC MOSFETs. Furthermore, the methods reported in [7] and [8] require additional external electronic circuitry.

In [9], the rate of change of the drain-source voltage ($dv_{ds}/dt$) across the device is directly utilized to detect a short-circuit fault. This method is based on the operating principle that during a hard-switching fault event, the device voltage remains near the dc bus voltage, yielding a negligible $dv_{ds}/dt$.

Conversely, during a normal turn-on transition, the voltage drops rapidly from the dc bus voltage to the low on-state conduction voltage, creating a distinct $dv_{ds}/dt$ profile. In [10], the DESAT short-circuit protection method, which is widely used for Si IGBTs, is applied to SiC MOSFETs. The device's on-state voltage is monitored and compared to a threshold voltage to detect a short-circuit fault. The method in [9] achieves faster short-circuit detection time by directly relying on $dv_{ds}/dt$, avoiding the blanking time that is usually present in [10]. However, unlike the DESAT method, the method in [9] fails to detect the fault under load conditions and uses the DESAT protection method as a second stage. In [12], a junction-temperature-based DESAT method is presented that adaptively sets the drain-source voltage threshold as a function of junction temperature to improve fault detection time. To minimize the blanking capacitance and achieve a faster fault response, the studies in [13] and [14] eliminate the physical blanking capacitor entirely, relying instead on the parasitic capacitance of the DESAT circuit components as the effective blanking capacitance.

Among all the short-circuit protection methodologies, current-sensing and gate-charge characteristic-based methods require significant electronic circuitry, which contributes to increased parasitics and switching losses. The DESAT method is advantageous because it can be easily integrated into commercially available gate drivers with minimal external components and has high noise immunity. Hence, the DESAT short-circuit protection method is widely adoptable for SiC MOSFETs [1], [2], [12], [13], [14], and [15]. Despite these advantages, [1], [2], [13], and [14] highlight the possible susceptibility of the DESAT circuit to false fault detection due to noise introduced by the $dv_{ds}/dt$ transitions, high-frequency oscillations in the drain-source voltage, and the reverse recovery charge of the DESAT diode.

While the effects of high-frequency device voltage oscillations during turn-off and the reverse recovery charge of the DESAT diode on the DESAT circuit are analyzed in detail in [13], [14], a quantitative analysis of the effects of $dv_{ds}/dt$ on the DESAT circuit remains limited in the literature. This study is essential because the rapid $dv_{ds}/dt$ voltage transition from the on-state to the full dc bus voltage occurs during every switching period in any typical power electronic converter. Because this voltage transient occurs prior to the high-frequency device voltage oscillations, its independent impact on the DESAT circuit is significant and cannot be neglected. Although internal deglitch filters in commercially available gate driver ICs provide noise immunity, a quantitative analysis of $dv_{ds}/dt$ effects remains imperative. Such an analysis reveals the fundamental trade-off between noise induced due to $dv_{ds}/dt$ and fault-detection speed, enabling the optimal selection of DESAT circuit parameters to achieve both high noise suppression and short-circuit detection times suitable for SiC MOSFETs.

Furthermore, while current-sensing and voltage-sensing methods are compared in detail in [1] and [2], a systematic comparison detailing the performance and design trade-offs among the architectural variants of DESAT protection circuits also remains limited in the literature. Therefore, the main objective of this article is to provide designers with a quantitative guideline for selecting the appropriate DESAT circuit variant and corresponding component values, based on the analysis and performance evaluation presented here, tailored to SiC MOSFETs. The major contributions of this article, which is an extended version of [16] are listed as follows

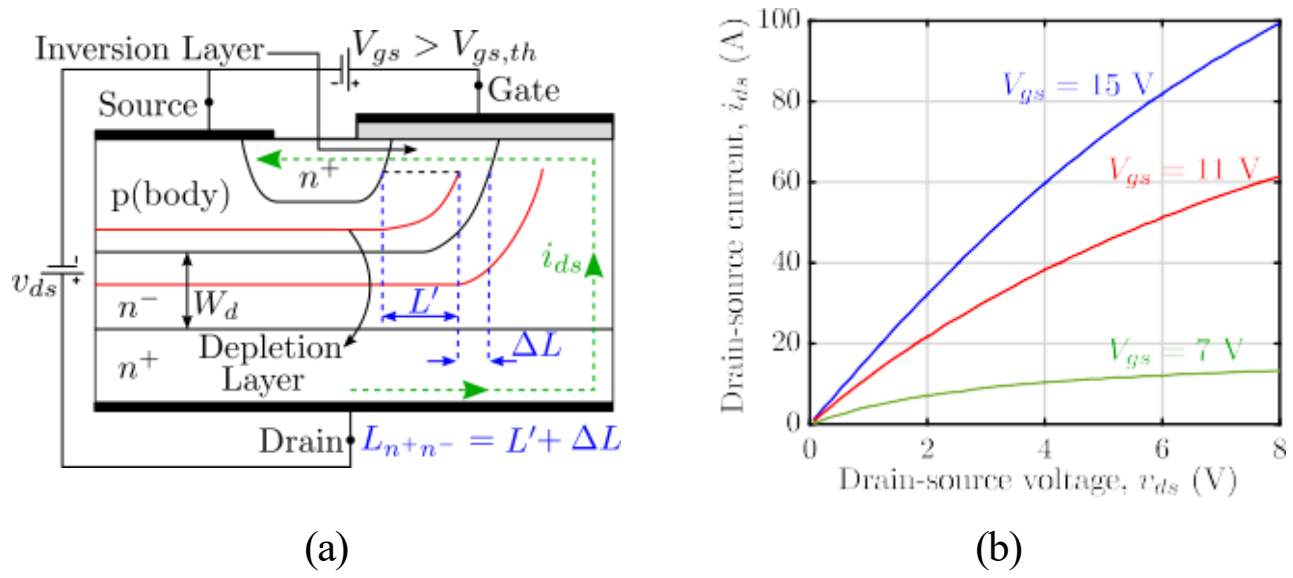

Fig. 1. (a) Cross section of SiC MOSFET showing the short channel effect during forward conduction, and (b) output characteristics of SiC MOSFET C3M0060065K (DUT).

1) This article evaluates three DESAT circuit configurations, providing design guidelines and operational limits for SiC MOSFET applications.
2) While previous studies [1], [2] have highlighted that $dv_{ds}/dt$ transients during SiC MOSFET turn-off induce displacement currents into the DESAT circuit, which momentarily charge the blanking capacitor above the threshold voltage and cause false short-circuit tripping, which remains largely qualitative. To address this, the article presents a quantitative analysis of the induced peak voltage across the blanking capacitor due to the $dv_{ds}/dt$ transients during normal turn-off conditions.
3) The design methodology for the DESAT methods suitable for SiC MOSFET are established from analyzing the $dv_{ds}/dt$ induced effects, which aids in the proper selection of the DESAT circuit component values which maintains the peak induced voltage across blanking capacitor within a window of $1-2$ V above the threshold voltage of the gate driver IC. This protects the gate driver IC, balances the induced noise due to $dv_{ds}/dt$ during the turn-off transient, and provides short-circuit detection time suitable for SiC MOSFETs.

## II. Short Circuit Behavior and Evaluation of Short Circuit Withstand Time

### A. Short Channel Effects on Output Characteristics of SiC MOSFET

Due to the higher breakdown field strength of SiC material, SiC MOSFETs feature a significantly thinner $n^-$ drift region ($W_d$) with a higher doping concentration ($N_d$) compared to Si counterparts, leading to a much lower specific on-state resistance [17]. However, the higher doping and smaller geometry lead to short-channel effects [18]. During saturation under high $v_{ds}$, the depletion region extends into the inversion layer by a length $\Delta L$ as shown in Fig. 1(a), reducing the effective inversion layer length to $L' = L_{n^+n^-} - \Delta L$. This channel-length modulation causes the gate-source threshold voltage $V_{gs,th}(v_{ds}, T_j)$ to decrease as $v_{ds}$ increases. Consequently, the saturation drain current is expressed as [19]

$$i_{ds} = \frac{1}{2}\mu_n C_{ox}\frac{W}{L'}\left[v_{gs} - V_{gs,\text{th}}\left(v_{ds}, T_j\right)\right]^2 . \tag{1}$$

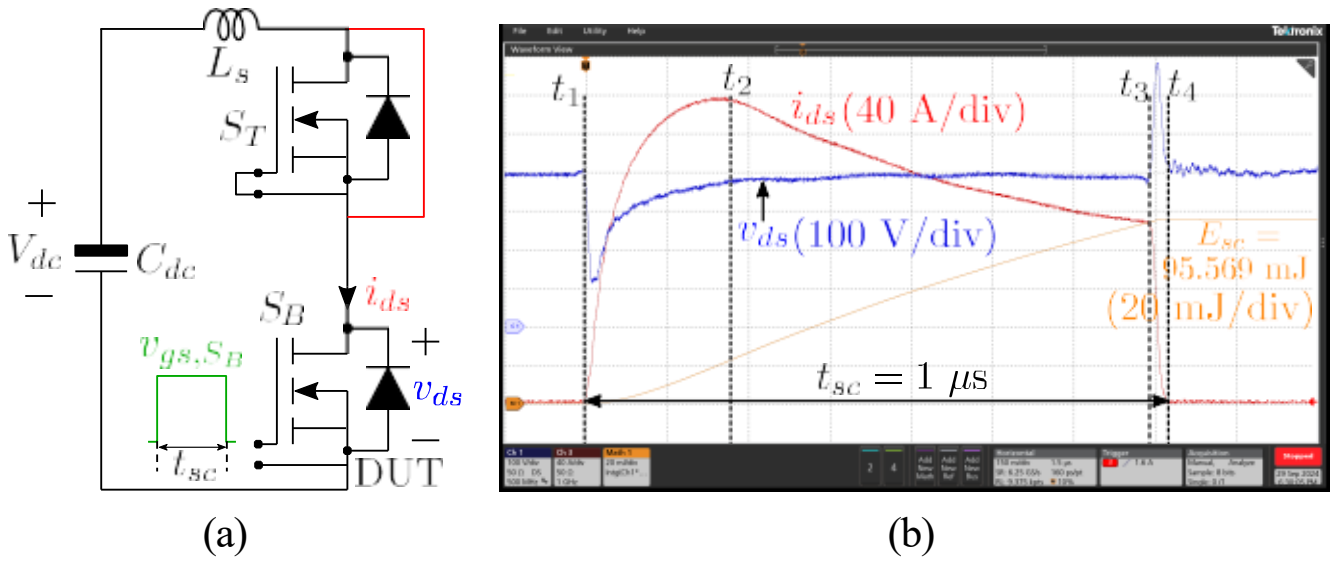

Fig. 2. (a) The test circuit configuration of non-destructive single-pulse short circuit test, (b) Experimental results showing $i_{ds}$ and $v_{ds}$ of the DUT [20] for a 1 $\mu$s short circuit pulse at dc bus voltage of 400 V, $v_{gs,S_B}$ = 15 V/ −4 V, and case temperature of $T_c$ = 25 °C.

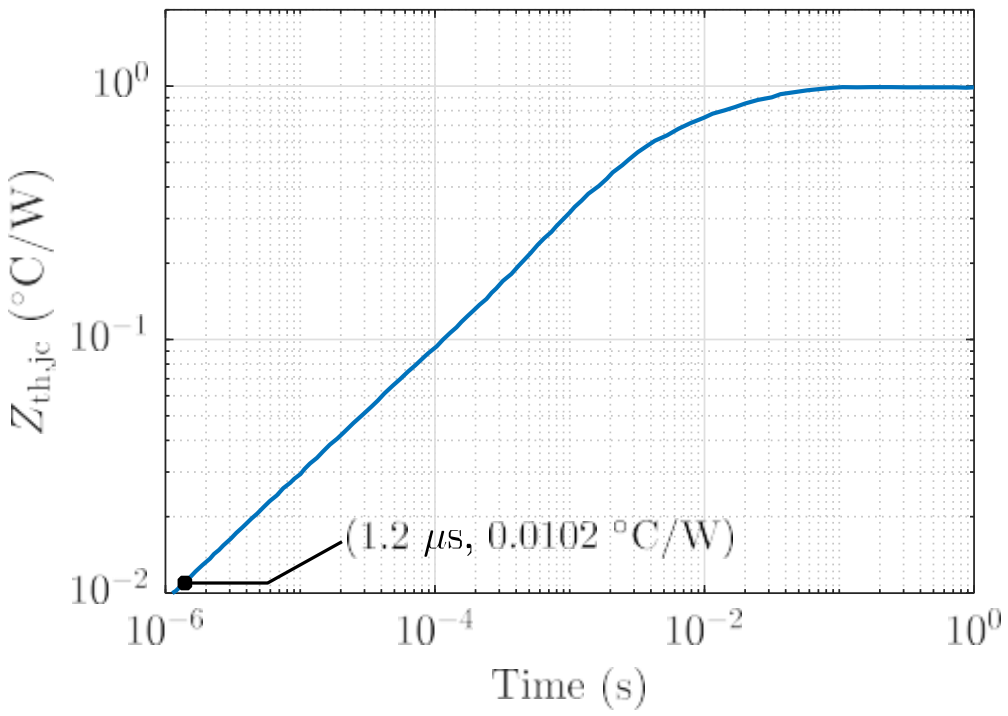

Fig. 3. Digitized single-pulse transient thermal impedance curve of C3M0060065K [20] used to determine the SCWT.

Because $V_{gs,th}$ decreases and $L'$ shortens with $v_{ds}$, $i_{ds}$ increases almost linearly with rising $v_{ds}$ in the active region, rather than exhibiting the flat saturation curves typical of Si IGBTs, as shown in Fig. 1(b). As a result, the peak short-circuit current of SiC MOSFETs can reach up to ten times the nominal current rating [1], demanding significantly faster short-circuit protection compared to Si IGBTs.

### B. Short Circuit Behavior of SiC MOSFET

To understand the short circuit behavior of SiC MOSFET, the device under test (DUT) [20] is subjected to a short circuit for 1 $\mu$s by disabling the short-circuit protection in the gate-driver. The circuit configuration with the bottom device ($S_B$) as the DUT and with the top device ($S_T$) shorted externally, as shown in Fig. 2(a), is used for this non-destructive single-pulse short-circuit test. The test conditions are at dc bus voltage of $V_{dc}$ = 400 V, gate source voltage of $v_{gs,S_B}$ = 15 V/ − 4 V, gate resistance $R_g$ = 5 Ω, and at case temperature $T_c$ = 25 °C. The experimental waveforms of current ($i_{ds}$) through and voltage ($v_{ds}$) across the DUT are shown in Fig. 2(b) for the device [20]. The total short circuit pulse duration ($t_{sc}$) is divided into three intervals. The mechanisms underlying the waveforms in each interval are discussed in detail below.

*1) Interval $t_1 \sim t_2$:* When the DUT is turned on into a short-circuit fault, it quickly transitions from the cut-off region to the active region as the gate voltage increases above the threshold voltage, with a high rate of change in current ($di_{ds}/dt$) through it. The $di_{ds}/dt$ depends on the power loop stray inductance ($L_s$), gate resistance, transconductance gain, $v_{gs}$ and $V_{dc}$ [1]. It is measured in this interval as 2.15 A/ns. The voltage across the DUT suddenly drops from $V_{dc}$ due to $di_{ds}/dt$ in the $L_s$ given by

$$v_{ds} = V_{dc} - L_s\frac{di_{ds}}{dt} . \tag{2}$$

Further $i_{ds}$ continues to increase towards the saturation point corresponding to the applied $v_{gs}$. Also, $v_{ds}$ recovers from the dip to finally settle at $V_{dc}$.

*2) Interval $t_2 \sim t_3$:* The DUT now conducts in the active region with full $V_{dc}$ across it. This causes significant power loss and, consequently, a rapid rise in device junction temperature due to self-heating. If left unprotected, the DUT experiences thermal failure. At these elevated temperatures, lattice vibrations severely decrease the electron mobility ($\mu_n$) in the channel and drift region. Since both the channel and drift region's resistance are inversely related to $\mu_n$, the increase of the combined channel and drift-region resistance suppresses the short circuit current, and hence the negative $di_{ds}/dt$ slope is observed during this interval [21].

*3) Interval $t_3 \sim t_4$:* At $t_3$, the gate pulse for the DUT is removed, and $i_{ds}$ rapidly decreases with a negative slope. According to (2), a negative $di_{ds}/dt$ causes a voltage overshoot across the DUT as shown in Fig. 2(b).

These insights into the short-circuit behaviour provide key design metrics for developing effective short-circuit protection methods. The essential design criteria that must be satisfied are as follows:

(i) The short-circuit fault detection time (SCDT) must be shorter than the duration required for the transient junction temperature ($T_j$) to reach its thermal failure limit.
(ii) A turn-off procedure from short-circuit to protect the DUT from voltage breakdown due to voltage overshoot.

### C. Evaluation of Short Circuit Withstand Time

Evaluation of short circuit withstand time (SCWT) is necessary to determine the SCDT. This withstand time can be analytically estimated by utilizing the single pulse junction-case transient thermal impedance ($Z_{\text{th,jc}}$) curve along with the maximum junction temperature limit that triggers thermal failure. For instance, literature [21] reports that a second-generation SiC MOSFET reaches a maximum $T_j$ of approximately 1000 °C immediately prior to catastrophic thermal failure. Although this article utilizes a third-generation SiC MOSFET from the same manufacturer, the same device failure temperature of 1000 °C is assumed as baseline in evaluating SCWT to avoid the destructive single pulse short-circuit test.

The average short circuit power loss ($P_{sc}$) dissipated within the device is calculated from the experimental short circuit

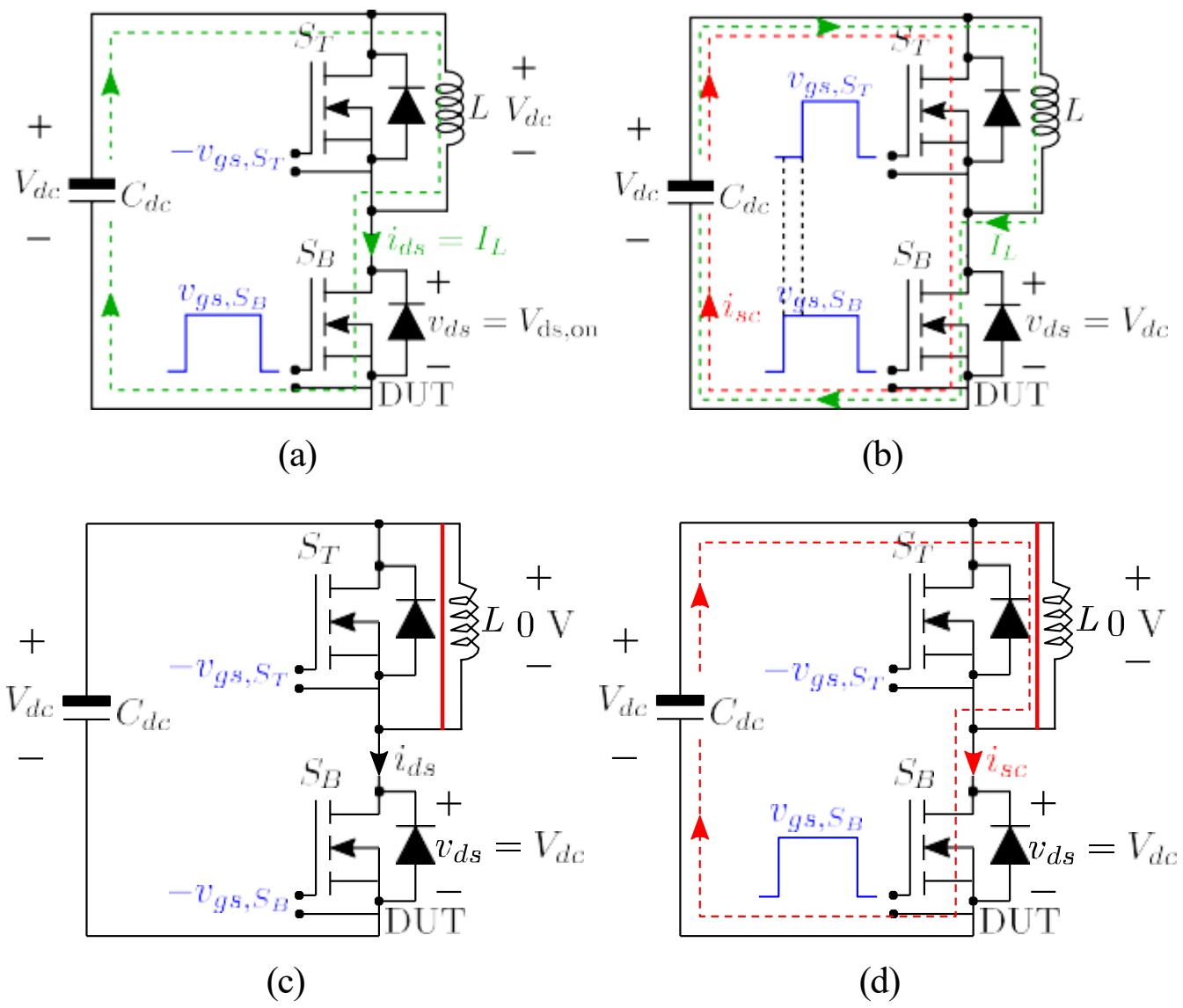

Fig. 4. Equivalent circuits (a) under steady state operation before FUL, (b) when DUT is subjected to FUL, (c) circuit condition before HSF, and (d) when DUT is subjected to HSF, $i_{sc}$ is the short-circuit current.

energy ($E_{sc}$) as shown in Fig. 2(b) over a $t_{sc}$ = 1 $\mu$s pulse duration as

$$P_{sc} = \frac{E_{sc}}{t_{sc}}, \quad (3)$$

which is calculated as 95.57 kW for the device [20]. Assuming an initial case temperature of $T_c$ = 25 °C, the maximum allowable temperature rise to reach the failure boundary is $\Delta T_{j,\max}$ = 975 °C. The critical short-circuit energy and SCWT being nearly independent of case temperature [21], the assumption on initial case temperature $T_c$ = 25 °C is valid. The $Z_{\mathrm{th,jc}}$ required to induce this thermal failure is expressed as

$$Z_{\mathrm{th,jc}} = \frac{\Delta T_{j,\max}}{P_{sc}}. \quad (4)$$

This is calculated as 0.0102 °C/W. The corresponding SCWT can be obtained using this calculated $Z_{\mathrm{th,jc}}$ from the transient thermal impedance curve shown in Fig. 3 as 1.2 $\mu$s. This analytically estimated SCWT represents the theoretical boundary for catastrophic thermal failure, rather than a permissible operating threshold.

Therefore, considering a derating factor of 0.5 applied to SCWT, the target SCDT is determined to be less than or equal to 600 ns. The SCDT should necessarily be less than the SCWT so that the device turns off before its junction temperature reaches the thermal failure limit. Therefore, short-circuit protection should be designed to detect various faults within the targeted SCDT.

## III. Types of Short Circuit Faults

Typically, short-circuit faults in a phase leg configuration occur due to the overlap of the gating pulses of the top device $S_T$ and the bottom device $S_B$ or when any of $S_T$ and $S_B$ is turned on into an existing short-circuit of the load. The overlap can happen due to incorrect gating logic or human error. Based on the steady-state condition prior to the fault, short-circuit faults are classified into fault under load and hard switching fault [22]. These two types of faults can be emulated on a double pulse test setup, and the effectiveness of the designed short circuit protection circuitry can be evaluated.

*1) Fault Under Load (FUL):* Prior to a fault condition, the DUT $S_B$ is conducting a steady state load current $I_L$ with drain-source voltage/device voltage $v_{ds}$ at its on-state voltage drop $V_{\mathrm{ds,on}}$. The complementary device $S_T$ is blocking the $V_{dc}$ as shown in Fig. 4(a). Under such steady-state conditions, a fault occurs if $S_T$ is falsely turned on. This leads to a short-circuit of the charged dc bus, resulting in a rapid increase in short-circuit current $i_{sc}$ in the DUT, Fig. 4(b).

*2) Hard Switching Fault (HSF):* DUT is subjected to HSF when it is turned on while an existing short-circuit is present. Prior to this fault condition, a short-circuit appears across the load inductor ($L$). This may be due to a short circuit in the device $S_T$ or in $L$. $S_B$ is in off state, $V_{dc}$ appears across it as shown in Fig. 4(c). Under such conditions, a fault occurs if $S_B$ turns on. This shorts the charged dc bus, which rapidly increases $i_{sc}$ through $S_B$. The voltage across $S_B$ remains at $V_{dc}$ to satisfy KVL as shown in Fig. 4(d).

Due to the zero initial voltage across $C_{\mathrm{blk}}$, the SCDT for an HSF event is inherently longer than that of a FUL. Consequently, HSF represents the worst-case condition for evaluating the SCDT. So, different types of DESAT short-circuit protection methods are analyzed only under HSF conditions in the subsequent sections.

TABLE I
Few Commercial Gate Driver ICs Based on the Source for Charging the DESAT Capacitor

| Gate Drives ICs | Source for Charging Blanking Capacitor | Category |
|---|---|---|
| ADuM4146 | Internal current source (500 $\mu$A) | Type-1 |
| NXP MC33GD3160 | Internal current source (250 $\mu$A - 1 mA) | Type-1 |
| ADuM4135 | Internal current source (537 $\mu$A) | Type-1 |
| UCC21750 | Internal current source (500 $\mu$A) | Type-1 |
| ACPL337J [23] | Internal current source (1 mA) | Type-1 |
| UCC21710 | External voltage source | Type-2 |
| UCC21732 | External voltage source | Type-2 |
| UCC21736 | External voltage source | Type-2 |

## IV. DESAT Short Circuit Protection Method

The DESAT short-circuit protection method can be implemented with various circuit configurations depending on the specific gate driver integrated chip (IC) selected. As summarized in Table I, these implementations are classified into two primary categories based on the source utilized for charging the blanking capacitor: type-1, which relies only on a fixed internal current source integrated within the IC, and type-2, which uses an external circuit configuration to charge the blanking capacitor.

Further sections systematically evaluate these implementation methodologies and address the distinct challenges they present when paired with fast-switching SiC MOSFETs. First, the limitations that cause the conventional DESAT protection circuit implemented with a type-1 IC to fall short of protection requirements are analyzed. Next, a specific circuit modification reported in the literature to upgrade type-1 for SiC compatibility is discussed. Finally, the DESAT protection circuit implemented with type-2 gate drivers is detailed.

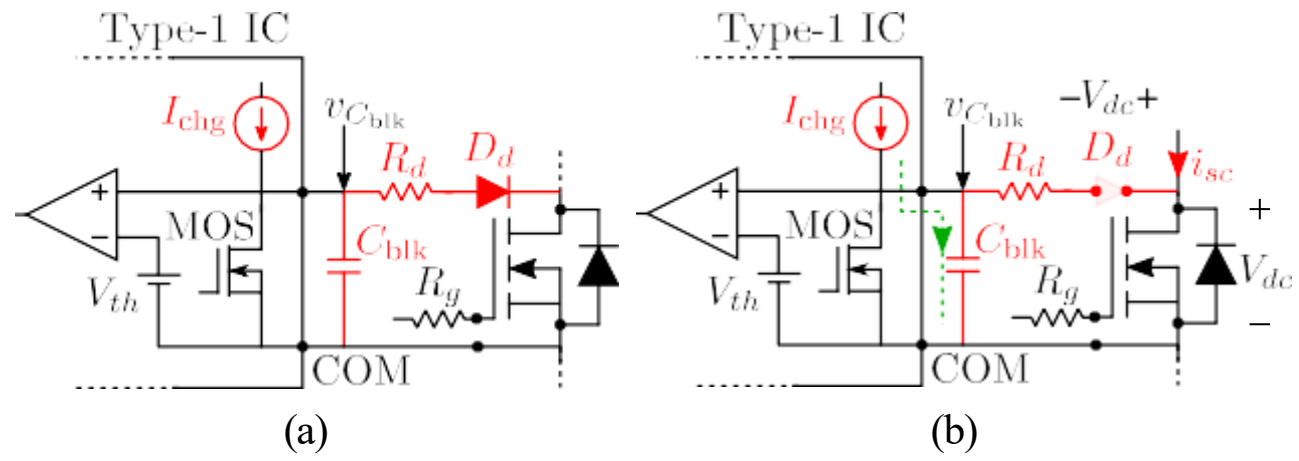

Fig. 5. (a) The DESAT protection circuit with type-1 gate driver IC, and (b) equivalent DESAT circuit operation during fault condition.

### A. DESAT Protection Method with Type-1 Gate Driver IC

The circuit diagram of the DESAT protection method utilizing a type-1 gate driver IC is illustrated in Fig. 5(a). The major components comprise an internal charging current source ($I_{\rm chg}$), a blanking capacitor ($C_{\rm blk}$), a high-voltage DESAT diode ($D_d$), and a series DESAT resistor ($R_d$). The diode $D_d$ is rated to block the $V_{dc}$, thereby isolating the low-voltage gate driver IC from the high-voltage power stage in certain condition detailed later in this section. The resistor $R_d$ limits excess current from the IC pin during voltage overshoots across the SiC MOSFET's anti-parallel body diode during turn-on transients. The detailed operating conditions of this DESAT protection circuit are discussed as follows:

*1) Operation During On-State of DUT:* During the on-state conduction of the SiC MOSFET, the diode $D_d$ is forward-biased, allowing the gate driver to continuously sense the device drain-source voltage $v_{ds}$. Under non-fault operating conditions, $I_{\rm chg}$ flows through the protection network and directly into the conducting power device. Consequently, the voltage across the blanking capacitor ($v_{C\,\rm blk}$) is clamped to a value well below the internal reference threshold voltage ($V_{th}$). This steady-state voltage across $C_{\rm blk}$ can be mathematically expressed as

$$V_{C_{\rm blk}} = I_{\rm chg}R_d + V_{f,D_d} + V_{\rm ds,on}, \tag{5}$$

where $V_{f,D_d}$ is the forward voltage drop of $D_d$, and $V_{ds,on}$ is the conduction voltage drop across the DUT.

*2) Operation When the DUT is in Off-State:* When the DUT is in the off-state, $v_{ds}$ rises to $V_{dc}$ in a typical one-leg configuration shown in Fig. 2(a). To prevent $I_{\rm chg}$ from charging the $C_{\rm blk}$ and causing a false fault trip during this period, an internal MOS within the gate driver IC activates, ideally clamping $v_{C_{\rm blk}}$ to 0 V. Concurrently, the high voltage at the drain node reverse-biases the diode $D_d$, which blocks $V_{dc}$ and isolates the low-voltage circuitry of the IC from the power stage.

*3) Operation During Fault Condition:* Under an HSF, when the DUT turns on from the off-state, the circuit is already shorted as shown in Fig. 4(c). So, $v_{ds}$ remains at the dc bus voltage $V_{dc}$, which keeps the diode $D_d$ in reverse-bias. However, since the device is in the on-state, the internal MOS is turned off by the gate driver. Consequently, $I_{\rm chg}$ begins to charge the $C_{\rm blk}$ linearly toward $V_{th}$, as illustrated in Fig. 5(b). Since the voltage across $C_{\rm blk}$ is initially 0 V during HSF, the time required ($t_{C_{\rm blk}}$) for the $v_{C_{\rm blk}}$ to reach $V_{th}$ is given by [1]

$$t_{C_{\rm blk}} = \frac{C_{\rm blk}V_{th}}{I_{\rm chg}}. \tag{6}$$

When $v_{C_{\rm blk}}$ exceeds the $V_{th}$, a short-circuit is detected and the internal comparator trips, communicating a fault signal across the galvanic isolation barrier to the primary side and also initiates the DUT's turn-off sequence.

TABLE II
NOMINAL PARAMETER VALUES USED FOR THE DESIGN ANALYSIS OF DESAT PROTECTION METHOD WITH TYPE-1 GATE DRIVER IC

| $V_{th}$ | $I_{\rm chg}$ | $t_{LEB}$ | $t_{GF}$ | $C_{\rm blk}$ |
|---|---|---|---|---|
| 7 V | 1 mA | 200 ns | 140 ns | 100 pF |

### B. Design Case Study with Type-1 Gate Driver IC

To illustrate the limitations of type-1 ICs for SiC MOSFET short-circuit protection, a practical design case is presented in this section. Typically, the range of $V_{th}$ and $I_{\rm chg}$ are 7 V − 9 V and 0.5 mA − 1 mA for type-1 ICs. Considering the lowest $V_{th}$ = 7 V, highest $I_{\rm chg}$ = 1 mA and a $C_{\rm blk}$ value of 100 pF as nominal, using (6) $t_{C\,\rm blk}$ is calculated as 700 ns. However, the SCDT is the cumulative sum of the $t_{C_{\rm blk}}$, the gate driver's internal hardware Leading-Edge Blanking delay ($t_{LEB}$), and the gate driver's internal deglitch filter delay ($t_{GF}$), expressed as

$$\text{SCDT} = t_{C_{\rm blk}} + t_{LEB} + t_{GF}. \tag{7}$$

By considering the typical values of $t_{LEB}$ = 200 ns and $t_{GF}$ = 140 ns, the total SCDT is calculated as 1.04 $\mu$s. The parameter values used in this analysis are tabulated in the Table. II. These values are considered by examining the datasheets of few commercially available type-1 gate driver ICs listed in the Table. I. This quantitative design analysis shows that the resulting SCDT of 1.04 $\mu$s exceeds the targeted design limit. Further, it leaves a narrow margin relative to the calculated SCWT of the DUT in Section II-C.

To meet the target design criteria of SCDT, (6) dictates that $t_{C_{\rm blk}}$ must be minimized. This can be achieved either by increasing $I_{\rm chg}$ or by reducing the value of $C_{\rm blk}$ to as low as 35 pF. This is also evident from [1], where a type-1 DESAT circuit is used with a very low value of $C_{\rm blk}$ = 24 pF to achieve SCDT of 500 ns. Such lower values of $C_{\rm blk}$ satisfy the SCDT requirement but could result in false fault detection [1], [2], and [13]. Further, the magnitude of $I_{\rm chg}$ is inherently constrained by the specific commercial gate driver IC selected. The problems associated with a lower value of $C_{\rm blk}$ are discussed as below.

### C. Problems Associated with Low Value of $C_{\rm blk}$

A lower value of $C_{\rm blk}$ improves the SCDT but reduces the noise immunity of the DESAT protection circuit to the fast-switching voltage transitions of the SiC MOSFET. The associated problems are detailed as below:

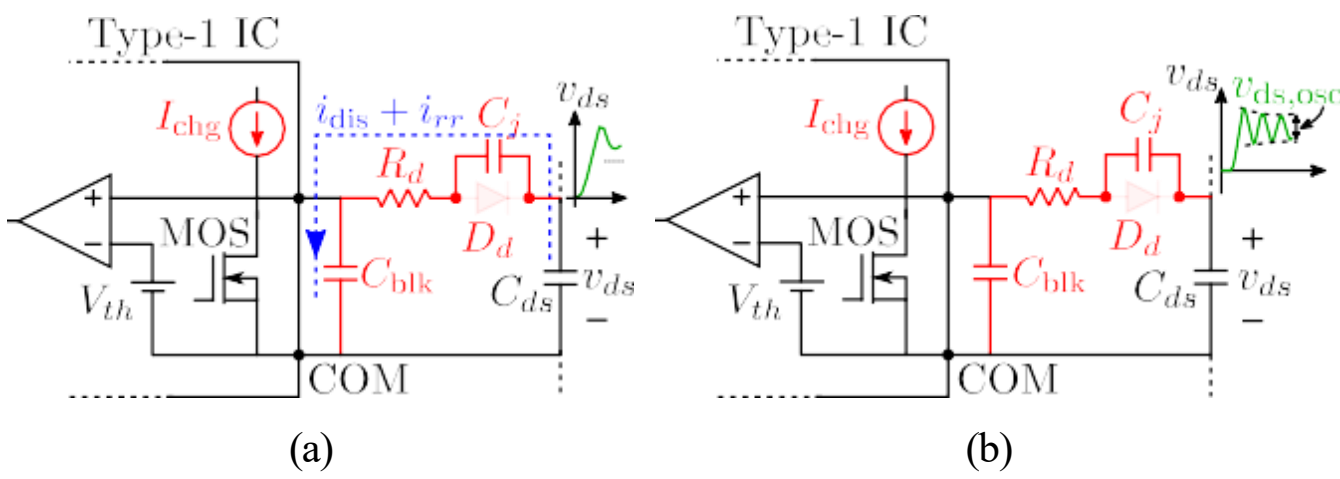

Fig. 6. DESAT protection circuit condition during normal turn-off of the DUT, where $C_{ds}$ is the drain-source capacitance of the DUT.

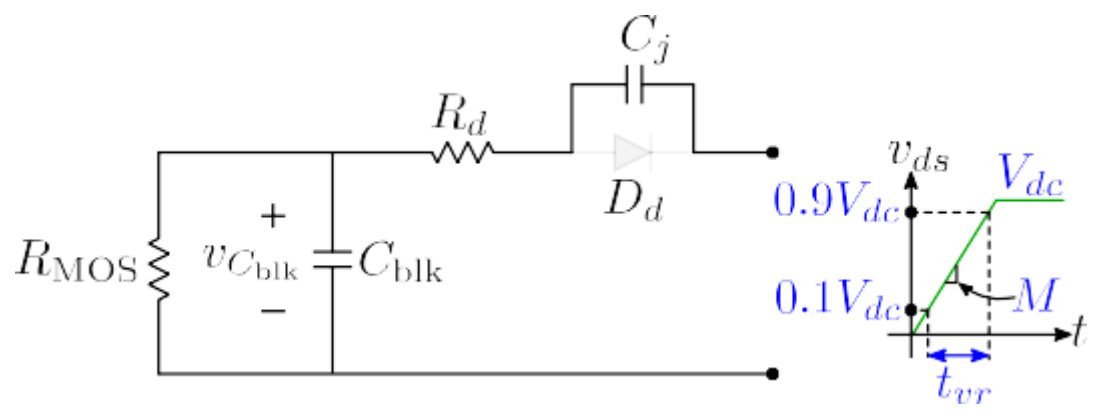

Fig. 7. Simplified equivalent circuit for analyzing the effect of $dv/dt$ of the DUT during turn-off transient on $v_{C_{blk}}$.

*1) $dv/dt$ induced false fault detection:* A low value of the $C_{blk}$, required to achieve SCDT suitable for SiC MOSFETs, can become comparable to the junction capacitance $C_j$ of the diode $D_d$. Under these conditions, during the normal turn-off transient of the DUT, the $dv/dt$ induced displacement current ($i_{dis} \approx C_j \frac{dv_{ds}}{dt}$) through $C_j$ together with the reverse recovery current ($i_{rr}$) of the diode $D_d$ can charge the $C_{blk}$ above $V_{th}$ [2]. This circuit condition is illustrated in Fig. 6(a), and the gate driver may report a false fault detection signal. Even though internal MOS clamps $v_{C_{blk}}$ when the PWM signal is low, this false detection can happen due to the finite resistance of the integrated MOS. Further, the peak magnitude of such a transient voltage across $C_{blk}$ can exceed the maximum voltage rating of the gate driver IC's DESAT pin [1] and may cause permanent damage to the IC or DESAT pin.

*2) High frequency oscillations in $v_{ds}$ induced false fault detection:* The second problem is that $C_j$ and $C_{blk}$ form a capacitor voltage-divider network at high frequency. The part of high-frequency oscillations in $v_{ds}$ during normal turn-off, as shown in Fig. 6(b), couples to $C_{blk}$ through $C_j$. The voltage across $C_{blk}$ can be analytically derived as

$$v'_{C_{blk}} = v_{ds,osc}\left(\frac{C_j}{C_j + C_{blk}}\right). \tag{8}$$

A lower value of $C_{blk}$ reduces the ratio $C_{blk}/C_j$, consequently charging $C_{blk}$ above $V_{th}$.

Literature [13] and [14] have analyzed the effects of $i_{rr}$ and high-frequency oscillations of device voltage during turn off on $v_{C_{blk}}$. However, the impact of $i_{dis}$ caused by $dv_{ds}/dt$ on $v_{C_{blk}}$ is not presented. Quantifying this displacement current effect on $v_{C_{blk}}$ is one of the contributions of this article and is analyzed in detail in the following sections.

## V. Impact of DESAT circuit parameters and $dv_{ds}/dt$ on $v_{C_{blk}}$

To satisfy the SCDT target established in Section II-C, minimizing the value of $C_{blk}$ is essential. Consequently, this section provides a quantitative sensitivity analysis of the behaviour of $v_{C_{blk}}$ as a function of the $dv/dt$ of the DUT, DESAT resistance $R_d$, capacitances $C_j$, and $C_{blk}$. The analysis assumes that $C_{blk}$ maintains an initial steady-state voltage ($V_{C_{blk,init}}$) given by (5) and the diode $D_d$ is in reverse bias condition, while the effect of $i_{rr}$ of the diode $D_d$ is neglected. The equivalent circuit used for the analysis is shown in Fig. 7, where $R_{MOS}$ is the finite ON-state resistance of the internal MOS. To analyze the effect on $v_{C_{blk}}$ with $dv/dt$ of $v_{ds}$, a relation between them is established in terms of the transfer function given by

$$V_{C_{blk}}(s) = M\left(\frac{A/B}{s\left(s^2 + (C/B)s + 1/B\right)}\right), \tag{9}$$

where $M$ represents the voltage gradient ($dv/dt$) of the $v_{ds}$ as shown in Fig. 7, $A = R_{MOS}C_j$, $B = R_{MOS}R_dC_{blk}C_j$, and $C = R_{MOS}C_{blk} + R_dC_j + R_{MOS}C_j$. By simplifying (9) and applying the inverse Laplace transform, the time domain expression for $v_{C_{blk}}$ is obtained as

$$v_{C_{blk}} = K\left(1 + \frac{\sigma_2}{\sigma_1 - \sigma_2}e^{-\sigma_1 t} - \frac{\sigma_1}{\sigma_1 - \sigma_2}e^{-\sigma_2 t}\right) + V_{C_{blk,init}}, \quad 0 \le t \le t_{vr} \tag{10}$$

where $K = MR_{MOS}C_j$, $\sigma_1$ and $\sigma_2$ are the real roots of $[s^2 + (C/B)s + 1/B]$, $t_{vr}$ is the voltage rise time of the DUT respectively. Once the active $dv/dt$ transition completes at $t \approx t_{vr}$, the diode $D_d$ enters a blocking state, isolating the low-voltage network. The current $i_{dis}$ stops, and the energy stored in $C_{blk}$ discharges naturally through $R_{MOS}$. After $t_{vr}$, $v_{C_{blk}}$ can be expressed as following

$$v_{C_{blk}} = V_i e^{-\tau/R_{MOS}C_{blk}}, \quad \tau > 0, \tag{11}$$

where $V_i$ is the voltage across $C_{blk}$ at $t = t_{vr}$, and $\tau = t - t_{vr}$ respectively. The peak voltage of $v_{C_{blk}}$ is extracted using (10), (11), and its sensitivity with the DESAT circuit parameters, including $dv/dt$, is analyzed further.

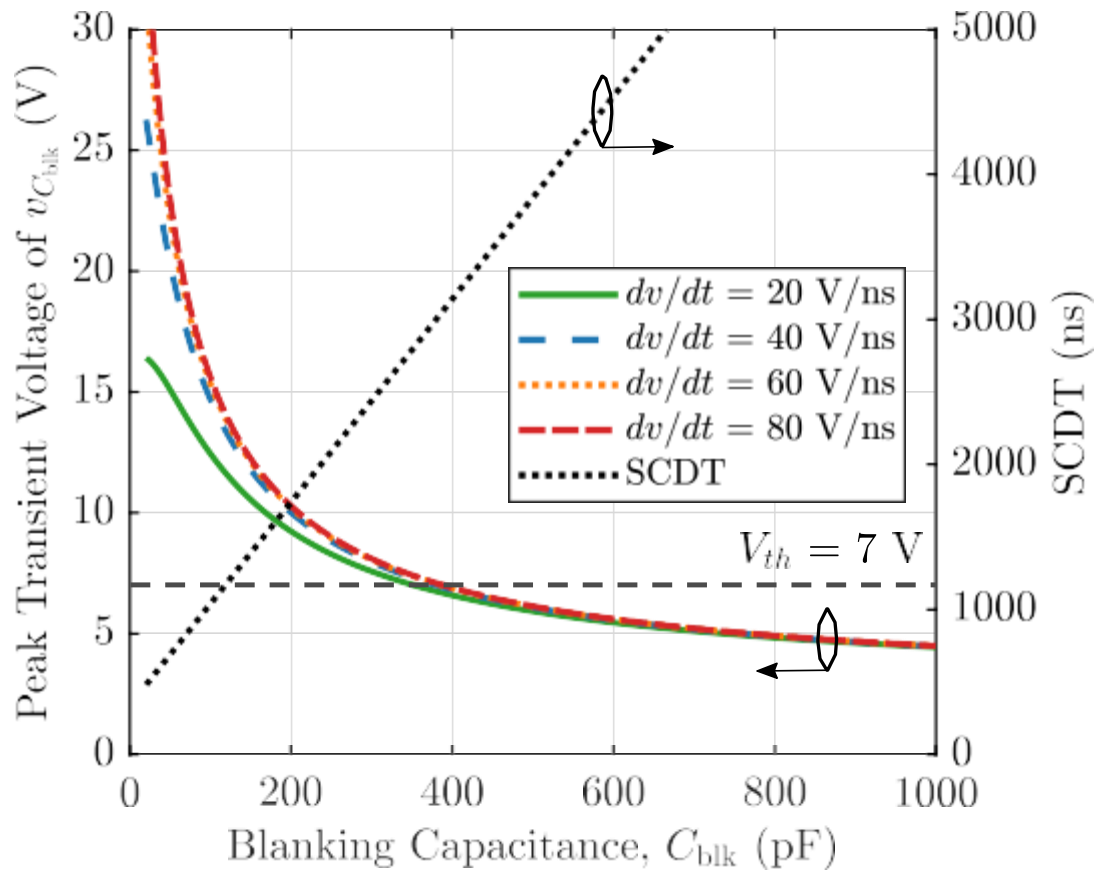

Fig. 8. Plot showing the sensitivity of the peak transient voltage of $v_{C_{blk}}$ and associated SCDT as a function of $C_{blk}$ at constant values of $R_d$ = 500 Ω, $C_j$ = 5 pF, and $R_{MOS}$ = 140 Ω.

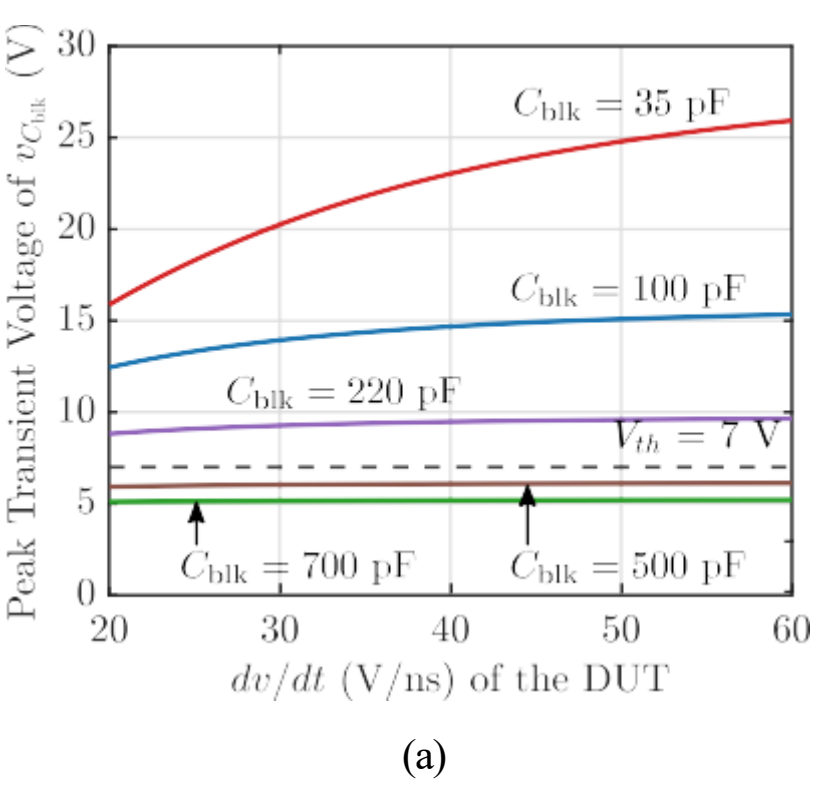

(a)

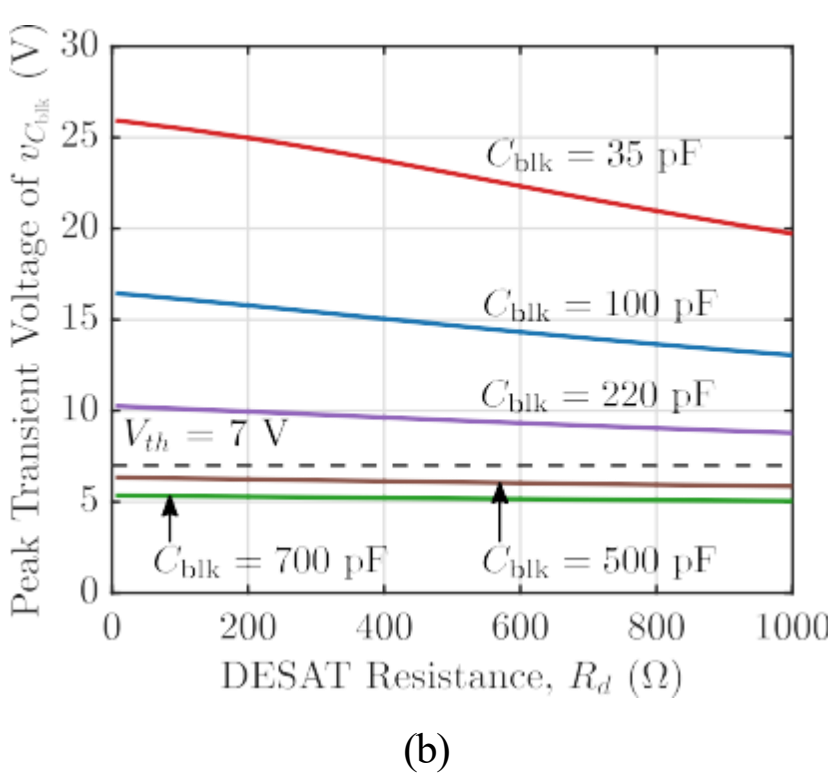

(b)

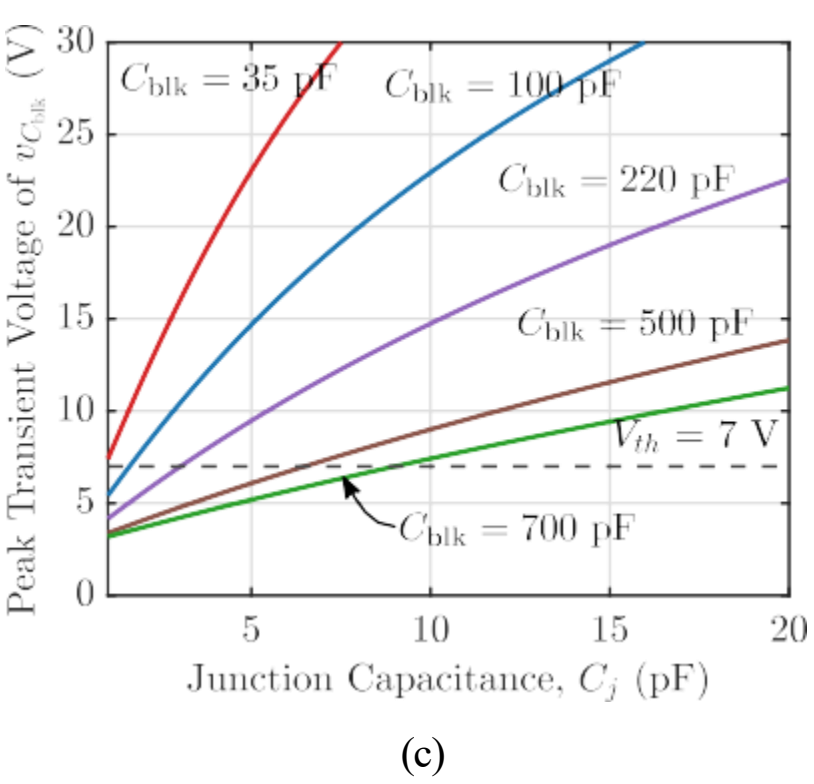

(c)

Fig. 9. Plots showing the sensitivity of the peak transient voltage of $v_{C_{blk}}$ at various values of $C_{blk}$ as a function of (a) $dv/dt$ of the DUT at constant $R_d$ = 500 Ω and $C_j$ = 5 pF, (b) DESAT resistance $R_d$ at constant $dv/dt$ = 40 V/ns and $C_j$ = 5 pF, and (c) junction capacitance $C_j$ at constant $dv/dt$ = 40 V/ns and $R_d$ = 500 Ω.

### *A. Variation of the Peak Transient Voltage of $v_{C_{blk}}$ with $C_{blk}$*

As shown in Fig. 8, a lower value of $C_{blk}$ = 35 pF causes the peak transient voltage of $v_{C_{blk}}$ to increase significantly higher than the $V_{th}$ across all the $dv/dt$ conditions. This excess voltage could trigger false fault detection [1], [2], [13] and [14]. Conversely, a higher value of $C_{blk} \geq 500$ pF attenuates the transient peak voltage below $V_{th}$ across the wide range of $dv/dt$ conditions, with a shortcoming of increased SCDT not suitable for SiC MOSFETs. This is also evident from the same plot, where SCDT increases linearly with $C_{blk}$. The SCDT is calculated as per (7). Furthermore, it is observed that for a given $dv/dt$, increasing $C_{blk}$ beyond 500 pF does not significantly change the attenuation of the peak transient voltage.

### *B. Variation of the Peak Transient Voltage of $v_{C_{blk}}$ with $dv/dt$*

Fig. 9(a) shows the effect of $dv/dt$ (20 V/ns − 60 V/ns) on the peak transient voltage of $v_{C_{blk}}$ at different values of $C_{blk}$. It can be observed that for values of $C_{blk} < 500$ pF, the voltage $v_{C_{blk}}$ is charged above $V_{th}$ across entire range of $dv/dt$ conditions. Conversely, larger values of $C_{blk} \geq 500$ pF successfully clamp the peak voltage safely below $V_{th}$ across the entire range of $dv/dt$, though they incur the SCDT penalties previously analyzed.

### *C. Variation of the Peak Transient Voltage of $v_{C_{blk}}$ with $R_d$*

Fig. 9(b) shows the variation of transient peak voltage as resistance $R_d$ is varied from 0 Ω to 1 kΩ. Although increasing $R_d$ reduces the peak transient voltage for the range 35 pF ≤ $C_{blk}$ < 500 pF, $v_{C_{blk}}$ still exceeds the $V_{th}$. Furthermore, increasing $R_d$ shows a very small effect on the peak voltage for the values $C_{blk} \geq 500$ pF.

### *D. Variation of the Peak Transient Voltage of $v_{C_{blk}}$ with $C_j$*

Fig. 9(c) shows the influence of $C_j$ varied from 1 pF to 20 pF on the peak voltage of $v_{C_{blk}}$. With low-capacitance values $C_{blk} \leq 100$ pF, a minor increase in $C_j$ forces the transient peak to rapidly increase above $V_{th}$. Further, the values of 2.5 pF < $C_j$ < 5 pF safely attenuate the transient peak voltage across the range 220 pF < $C_{blk} \leq 700$ pF.

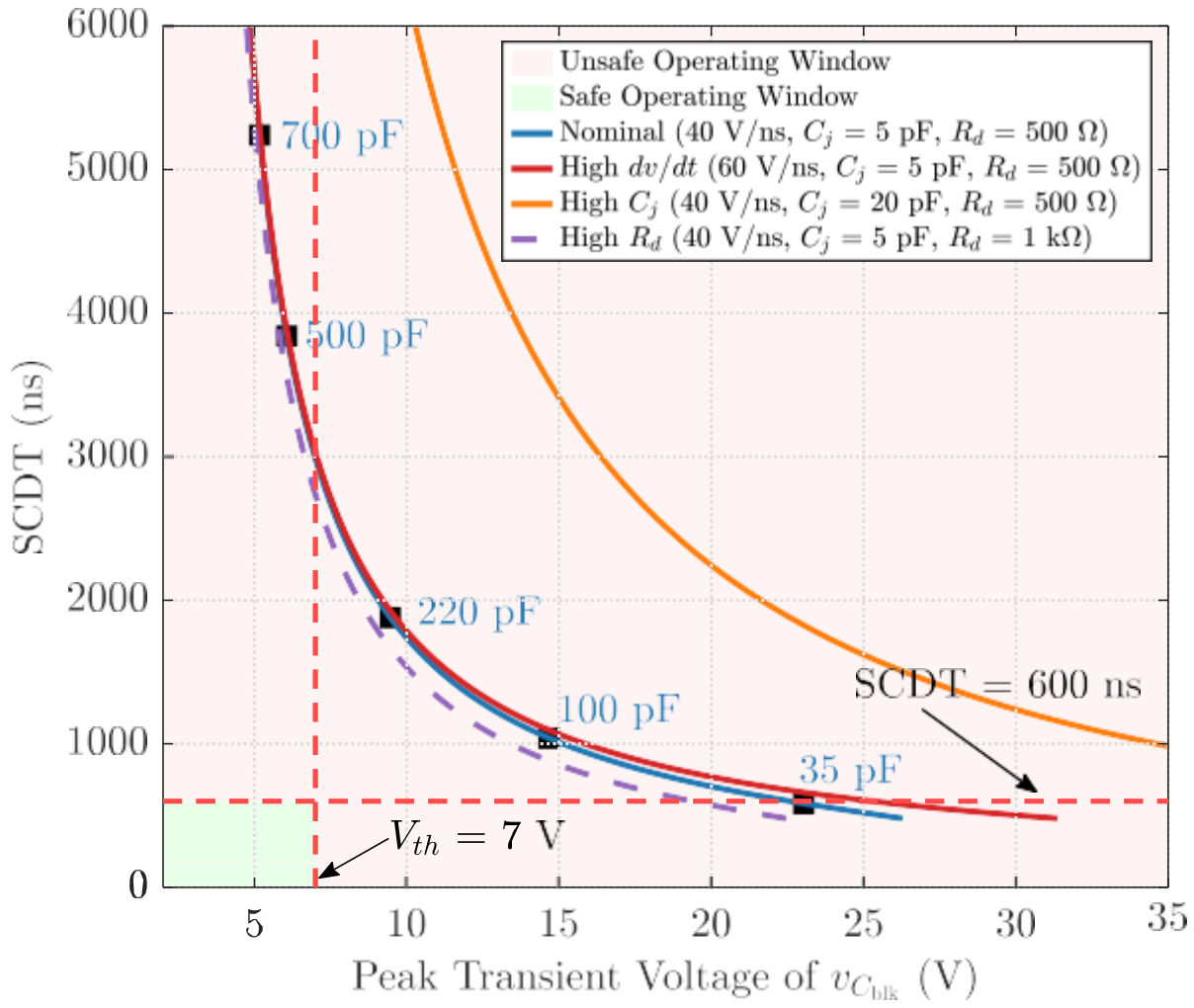

Fig. 10. Safe Operating Area (SOA) boundary curve for DESAT protection method with type-1 gate driver IC.

From the above analysis, for $C_{blk} \geq 500$ pF, the peak transient voltage of $v_{C_{blk}}$ becomes insensitive to variations in both $dv/dt$ and $R_d$, indicating that $C_{blk}$ dominates the noise immunity in this region. Although such large blanking capacitance values improve the noise immunity of the DESAT circuit shown in Fig. 5(a) against possible false fault detections, they fail to meet the SCDT requirements for SiC MOSFETs.

### *E. Safe Operating Area (SOA) Boundary Curve*

By combining the parametric data from Fig. 8 and Fig. 9, the safe operating area (SOA) boundary curves for the type-1 DESAT protection circuit (Fig. 5(a)) are constructed in Fig. 10. Here, the SOA is bounded by the $V_{th}$ limit and the target SCDT for SiC MOSFETs. As observed, the resulting SOA is extremely narrow, with none of the practical operating conditions satisfying both constraints simultaneously. Furthermore, a high $C_j$ = 20 pF shifts the boundary curves significantly to the right, away from the viable SOA. This occurs because a higher $C_j$ increases the magnitude of the $i_{\text{dis}}$ generated during the $dv/dt$ turn-off transient. It can also be observed that $C_j$ exerts a stronger influence on the peak transient voltage than the $dv/dt$ itself, as noticed by the substantially higher peak transient voltages observed with an increased $C_j$ compared to those under maximum $dv/dt$ conditions at a given $C_{\text{blk}}$.

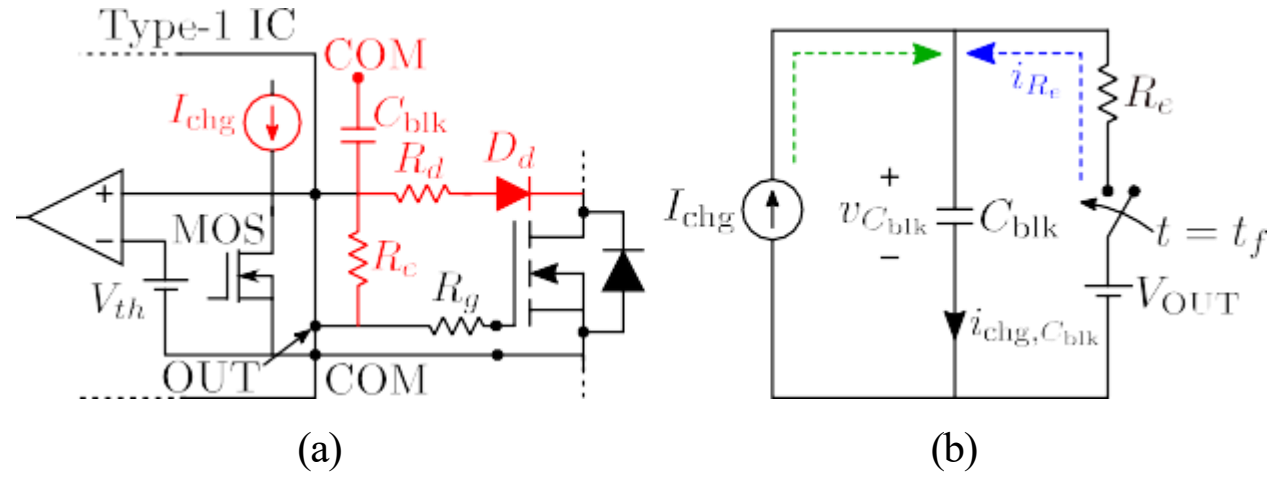

Fig. 11. (a) The DESAT protection method with type-1 gate driver IC including an additional resistor $R_e$, and (b) equivalent circuit when the DUT is switched into a fault condition.

Consequently, relying solely on *conventional DESAT protection circuit* parameters such as $R_d$, $C_j$, and $C_{\text{blk}}$ for SiC MOSFETs results in a *highly restrictive, marginal design space where optimizing for noise immunity directly compromises the SCDT*. Attenuating the transient peak voltage of $v_{C_{\text{blk}}}$ completely below $V_{th}$ requires a large $C_{\text{blk}}$. However, this large capacitance makes the peak voltage insensitive to $R_d$ and $C_j$ while introducing unacceptable detection delays.

Although internal deglitch filters can prevent false trips when transients exceed $V_{th}$, allowing unrestricted voltage spikes risks exceeding the maximum voltage rating of the IC's DESAT pin. *Therefore, this article aims at constraining the peak transient voltage of $v_{C_{\text{blk}}}$ to a controlled window of* $1-2$ V *above* $V_{th}$. As shown in Fig. 9 and 10, a capacitance range of 220 pF < $C_{\text{blk}}$ < 500 pF satisfies this limit while maintaining safe IC operation. To achieve SCDT suitable for SiC MOSFETs within this capacitance range, an additional degree of freedom was introduced in [24] to the DESAT circuit shown in Fig. 5(a), which is evaluated in the next section.

## VI. Improved DESAT Protection Circuit Architecture with Type-1 Gate Driver IC

As analyzed in the preceding section, a fundamental design conflict arises when reducing the value of $C_{\text{blk}}$ to satisfy the SCDT requirements of SiC MOSFETs. A smaller $C_{\text{blk}}$ inherently degrades the noise immunity during high $dv/dt$ transients. In [24], an additional resistor is connected between the OUT terminal and the DESAT pin of the gate driver IC. The circuit diagram is shown in Fig. 11(a). The gate driver OUT pin provides an extra charging current $i_{R_e}$ through an additional resistor $R_e$ to charge $C_{\text{blk}}$. This serves as an auxiliary charging path. The equivalent circuit when the DUT is switched into a fault condition at $t = t_f$ is shown in Fig. 11(b). The total charging current of the $C_{\text{blk}}$ is

$$i_{\text{chg},C_{\text{blk}}} = I_{\text{chg}} + i_{R_e}. \tag{12}$$

By applying transient network analysis to the equivalent circuit shown in Fig. 11(b), assuming a zero initial condition for $C_{\text{blk}}$, the updated $t_{C_{\text{blk}}}$ required for the capacitor $C_{\text{blk}}$ voltage to reach the threshold $V_{th}$ is derived as

$$t_{C_{\text{blk}}} = R_e C_{\text{blk}} \ln \frac{V_{\text{OUT}} + I_{\text{chg}} R_e}{V_{\text{OUT}} + I_{\text{chg}} R_e - V_{th}}. \tag{13}$$

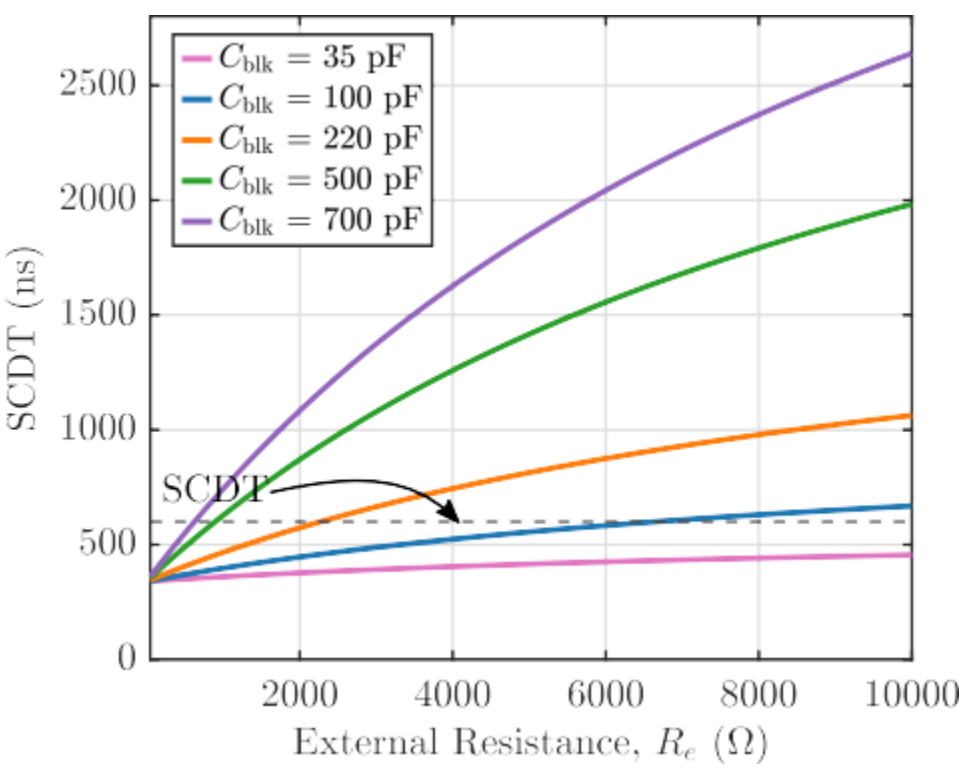

Fig. 12. Plot showing the variation of SCDT with external resistance $R_e$ across different values of $C_{\text{blk}}$.

Comparing the conventional blanking time expression in (6) with the modified relation in (13), it is evident that the resistor $R_e$ introduces a critical design degree of freedom. By properly sizing $R_e$, the charging time can be scaled down to meet the SCDT for SiC devices while keeping $C_{\text{blk}}$ large enough to suppress the transient peak voltage of $v_{C_{\text{blk}}}$ during normal turn-off switching transient. This helps to prevent the false-fault detection issues. Although literature [24] introduced an additional parameter $R_e$ to improve the SCDT. The detailed design guidelines for quantitatively selecting $R_e$ to achieve a suitable SCDT for SiC MOSFETs remain limited.

This article presents the selection procedure of $R_e$ from Fig. 12 plotted using (13). It quantitatively shows the variation of SCDT with $R_e$ across various values of $C_{\text{blk}}$. The plot shows that SCDT increases with $R_e$ due to increased resistance in the auxiliary charging path. For a low capacitance value $C_{\text{blk}}$ = 35 pF, the SCDT remains comfortably below the target value across nearly the entire range of $R_e$. But from the analysis established in the previous section, a smaller value of $C_{\text{blk}}$ fails to provide adequate noise immunity during high-$dv/dt$ turn-off events. For larger values of $C_{\text{blk}} \geq$ 500 pF chosen specifically to suppress transient peak voltage below $V_{th}$ as shown in the previous analysis, the SCDT becomes highly sensitive to the value of $R_e$. To maintain the SCDT within the target limit for the proposed range of 220 pF < $C_{\text{blk}}$ < 500 pF, the value of $R_e$ must be selected as $R_e \leq 500\ \Omega$, as shown in Fig. 12. Therefore, integrating an appropriately sized $R_e$ alongside 220 pF < $C_{\text{blk}}$ < 500 pF into the type-1 DESAT protection circuit effectively improves the SCDT suitable for SiC MOSFETs while constraining the peak voltage of $v_{C_{\text{blk}}}$ to $1-2$ V above $V_{th}$.

Despite the advantage of decoupling the noise immunity from short-circuit detection speed by adding $R_e$, the type-1 gate driver IC has a $V_{th}$ fixed in the range of $6-9$ V, which in general falls in the active region on the output characteristics of the DUT. This range of $V_{th}$ leads to higher SCDT [12]. Hence, the DESAT protection method with a type-2 gate driver IC is analyzed in detail in the following section.

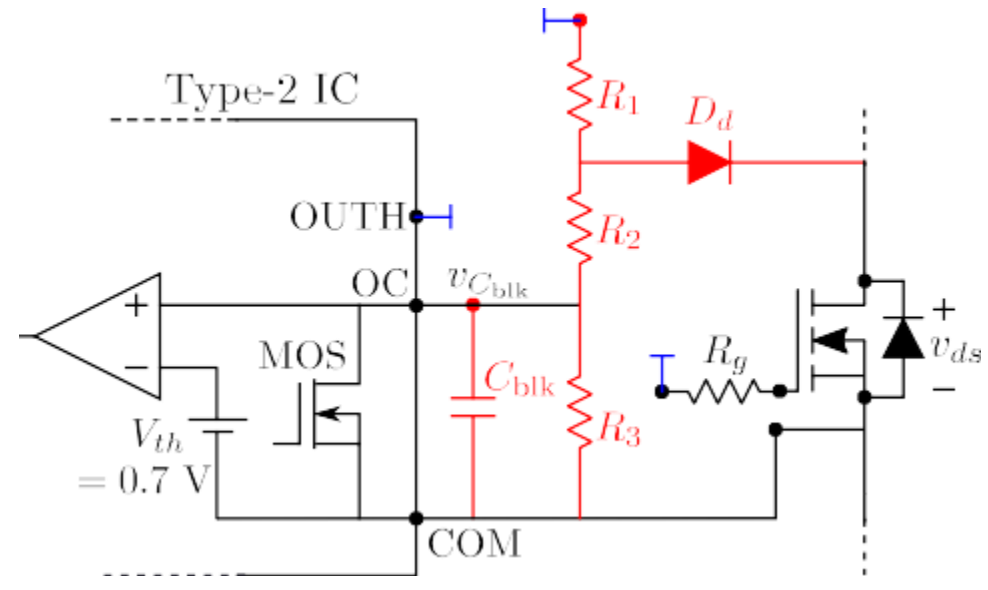

Fig. 13. The DESAT protection circuit with type-2 gate driver IC.

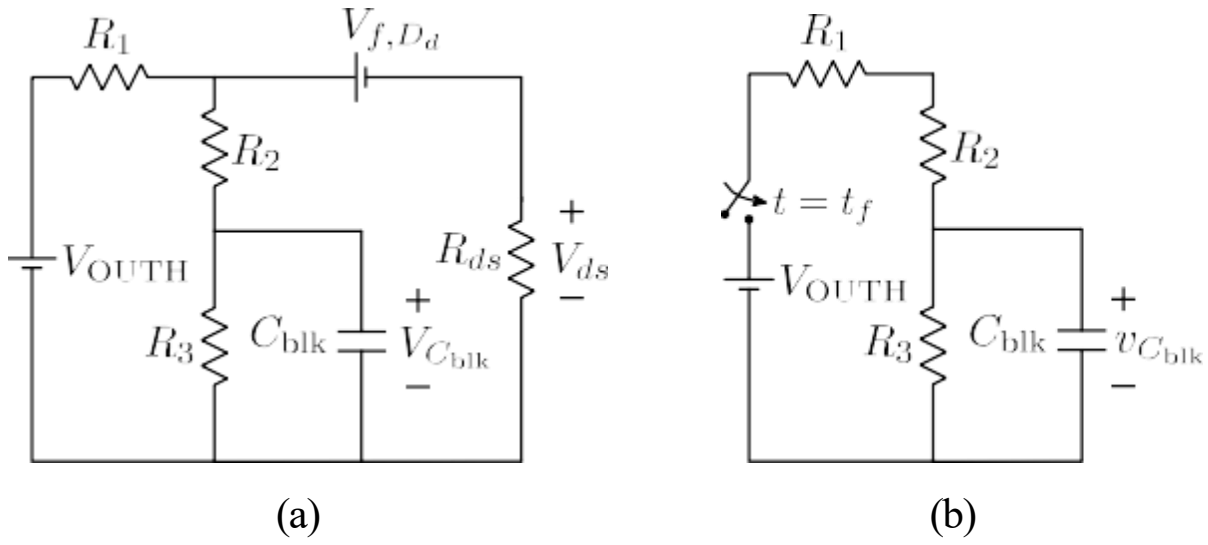

Fig. 14. (a) Steady state equivalent circuit when the DUT is conducting, and (b) equivalent circuit when the DUT is switched into a fault condition.

## VII. DESAT Protection Method with Type-2 Gate Driver IC

The DESAT protection circuit with type-2 gate driver IC is shown in Fig. 13. In this architecture, the conventional internal current source ($I_{\rm chg}$) is eliminated, and the capacitor $C_{\rm blk}$ is charged through an external resistive network directly connected to the OUTH pin of the gate driver IC. The internal $V_{th}$ is 0.7 V. The resistors $R_2$ and $R_3$ are used to set the drain-source short-circuit fault detection threshold voltage ($V_{\rm ds,det}$) along the output characteristic curves of the DUT, while this detection voltage is fixed in case of type-1 ICs as stated earlier. The steady state voltage across $C_{\rm blk}$ during normal conduction of the DUT is governed by the potential divider formed by $R_2$ and $R_3$ as shown in Fig. 14(a). This voltage can be expressed as follows

$$V_{C_{\rm blk}} = (V_{f,D_d} + V_{ds})\frac{R_3}{R_2 + R_3}. \tag{14}$$

When a fault condition forces $V_{C_{\rm blk}}$ to reach the internal comparator threshold, the protection circuit triggers. By substituting $V_{C_{\rm blk}} = V_{th}$ into (14), $V_{\rm ds,det}$ at which fault is detected is derived as

$$V_{\rm ds,det} = \frac{R_2 + R_3}{R_3} V_{th} - V_{f,D_d}. \tag{15}$$

When the DUT is switched into a fault at $t = t_f$, the time taken by the $C_{\rm blk}$ to charge till $V_{th}$ to issue a fault trip can be derived from the equivalent circuit shown in Fig. 14(b). It is expressed as

$$t_{C_{\rm blk}} = R_{eq} C_{\rm blk} \ln\left(1 - \frac{R_1 + R_2 + R_3}{R_3}\frac{V_{th}}{V_{\rm OUTH}}\right), \tag{16}$$

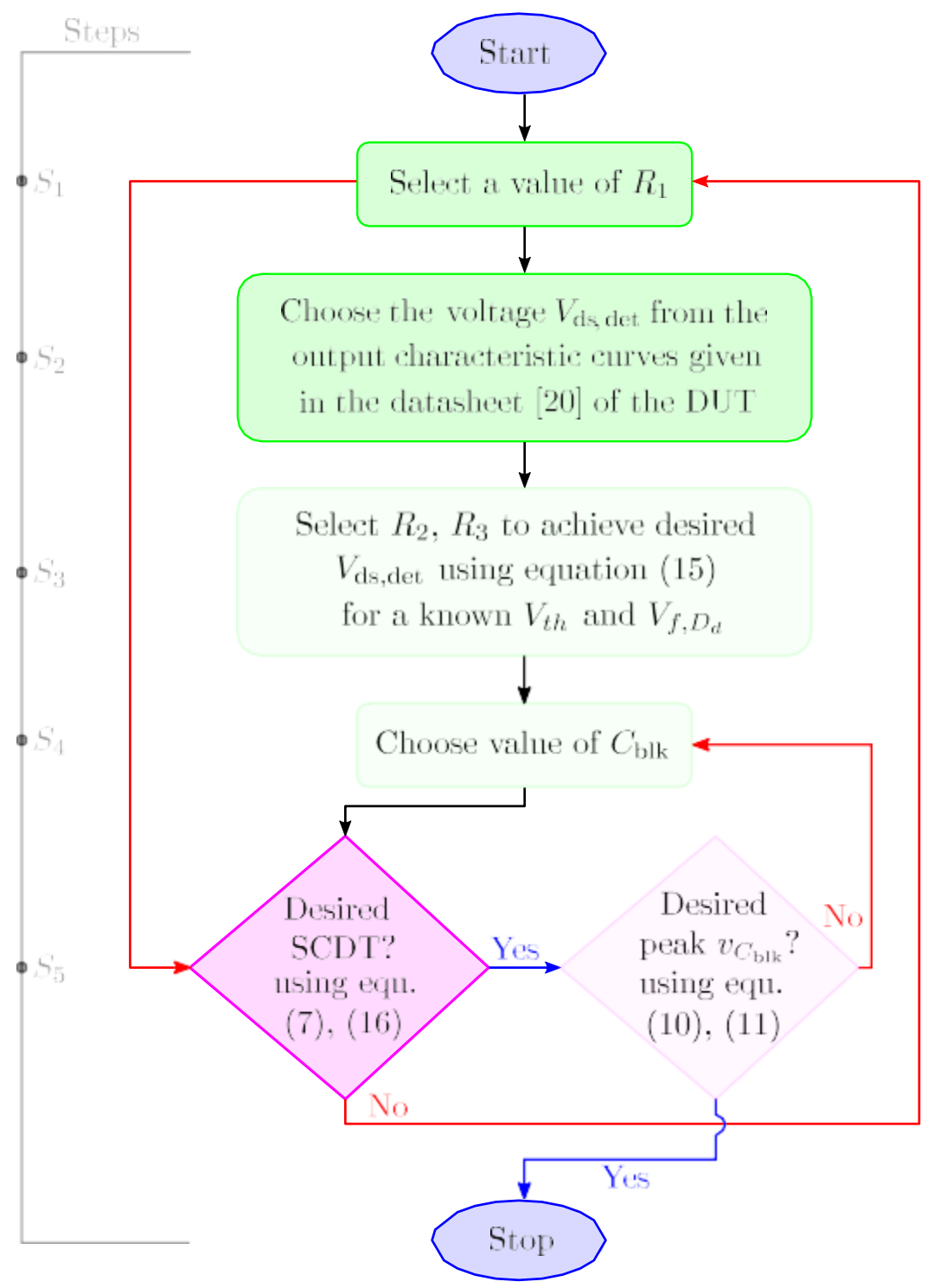

Fig. 15. Design flow chart to select the component values of the type-2 DESAT circuit.

where $R_{eq} = -\frac{(R_1 + R_2)R_3}{R_1 + R_2 + R_3}$. The values of $R_1$, $R_2$, and $R_3$ in (15) and (16) should be appropriately selected to achieve desired $V_{\rm ds,det}$ and $t_{C_{\rm blk}}$. Similarly, comparing the conventional blanking time expression in (6) with (16), it is evident that the resistors $R_1$, $R_2$, and $R_3$ provide three degrees of freedom to achieve desired SCDT with the attenuation limit on transient peak voltage of $v_{C_{\rm blk}}$.

Fig. 16(a) shows the variation of $V_{\rm ds,det}$ with resistances $R_2$ and $R_3$ at a fixed value of $R_1 = 600\,\Omega$ plotted using (15). The voltage $V_{\rm ds,det}$ is chosen for the worst case [19] such that the fault circuit should not issue a trip when the DUT is conducting 30 A (nearly continuous rated current of the DUT [20] at case temperature of 100 °C) at maximum $T_j$. Therefore, $R_2$ and $R_3$ are selected as 7 kΩ and 1450 Ω to set $V_{\rm ds,det} = 3.5$ V. For the known values of $R_2$ and $R_3$, $C_{\rm blk}$ should be selected to achieve desired SCDT and to attenuate the transient peak voltage of $v_{C_{\rm blk}}$ to 1 − 2 V window above $V_{th}$. Fig. 16(b) is plotted using (7) and (16), while Fig. 16(c) plots the transient peak voltage using the equivalent circuit similar to Fig. 7, and equations equivalent to (10), (11). This is because the equivalent circuit to analyze the impact of $dv_{ds}/dt$ on $v_{C_{\rm blk}}$ for the type-2 DESAT circuit is the same as that shown in Fig. 7 with different component values. From these plots, $C_{\rm blk} = 500$ pF is chosen, achieving a SCDT ≈ 580 ns while constraining the peak $v_{C_{\rm blk}}$ to approximately 1 V during normal turn-off of the DUT. Blanking capacitance values lower than 500 pF further improve the SCDT but significantly increase the magnitude of the transient peak voltage. A detailed design flow chart to select the component values of the type-2 DESAT circuit is shown in Fig. 15.

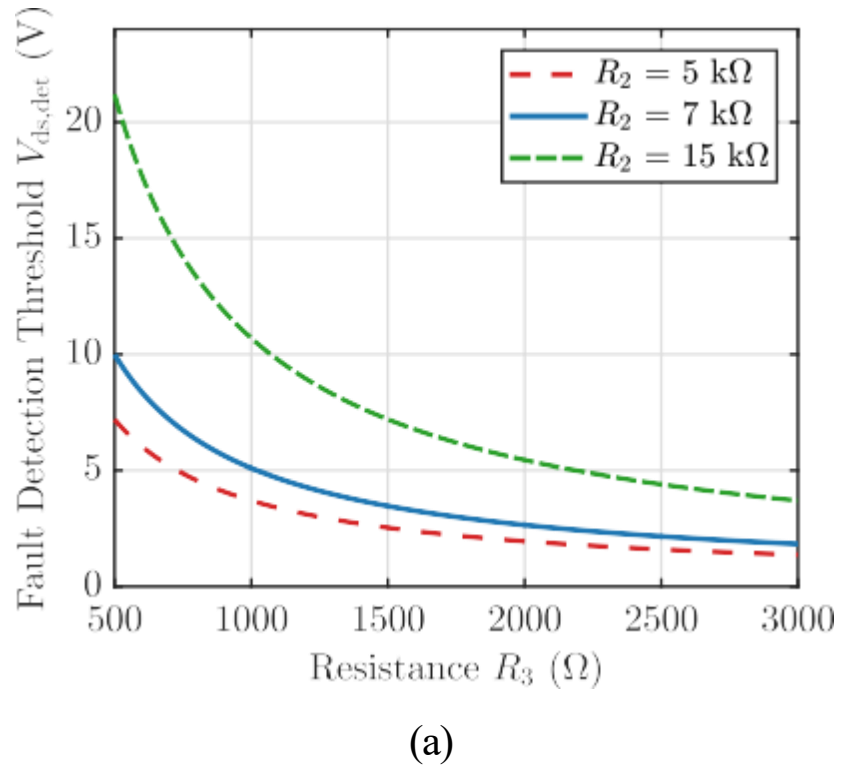

(a)

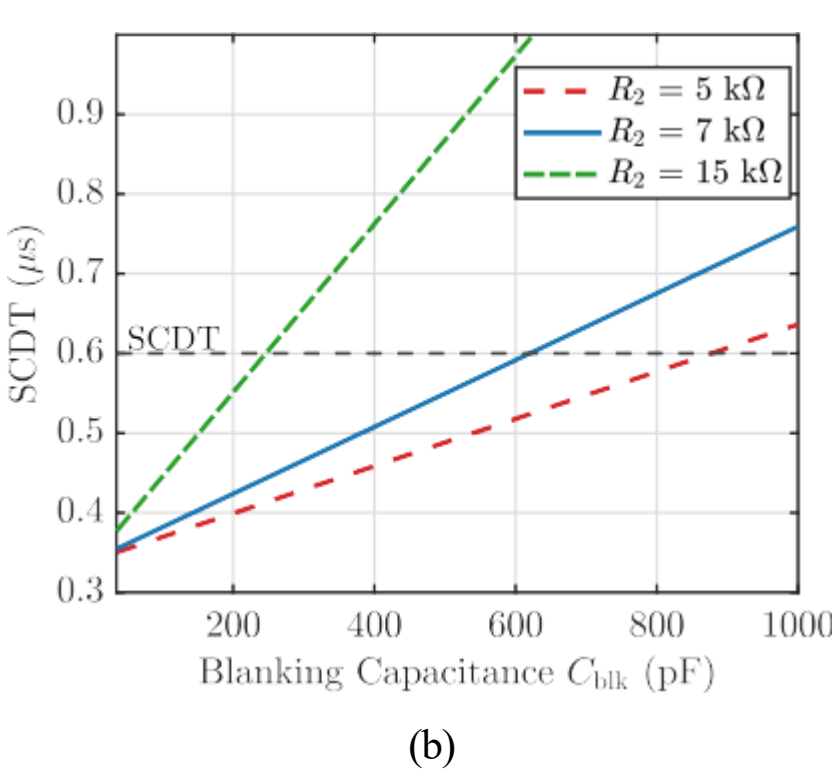

(b)

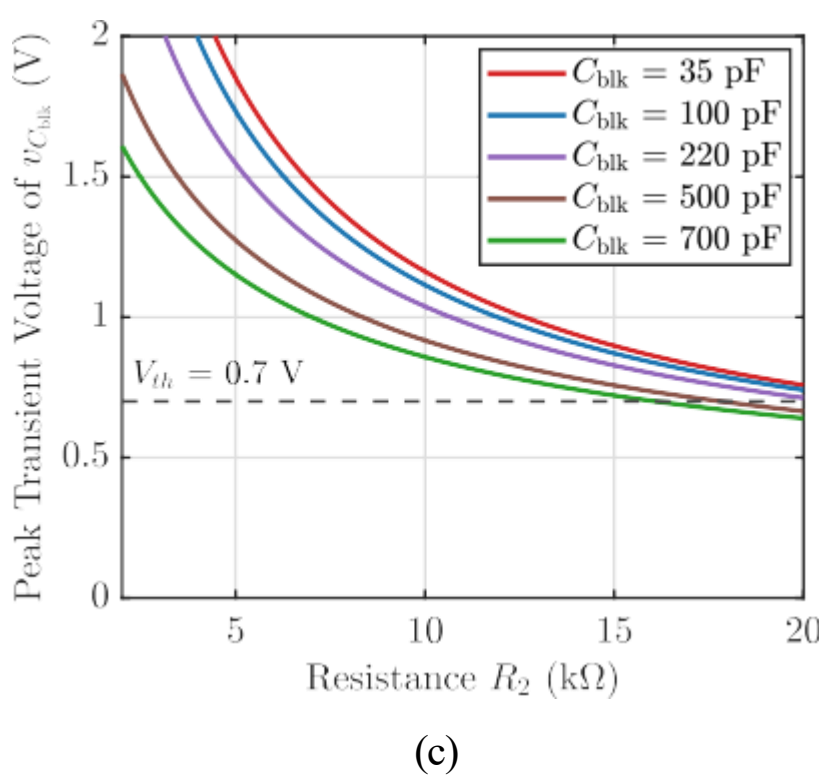

(c)

Fig. 16. Plot showing the variation of $V_{ds,det}$ varying with resistances $R_2$ and $R_3$ at $R_1 = 600\,\Omega$, (b) plot showing the variation of SCDT with $C_{blk}$ at different values of $R_2$ at $R_1 = 600\,\Omega$, $R_3 = 1450\,\Omega$, and (c) plot showing the peak transient voltage of $v_{C_{blk}}$ with different values of $R_2$ and $C_{blk}$ at $R_3 = 1450\,\Omega$, $dv/dt = 40\,$V/ns, and $C_j = 5\,$pF.

TABLE III
COMPREHENSIVE EVALUATION AND PARAMETRIC COMPARISON OF DESATURATION PROTECTION ARCHITECTURES FOR SIC MOSFET

| Protection Circuit Architecture | Target SCDT Achievability | $dv/dt$ Noise Immunity | Circuit Complexity | Fault Detection Threshold Voltage ($V_{ds,det}$) | Parameter to Decouple SCDT and Noise Immunity | Blanking Time ($t_{C_{blk}}$) |
|---|---|---|---|---|---|---|
| Conventional DESAT Fig. 5(a) | Tunable only with $C_{blk}$ | Low. Vulnerable to $dv/dt$ transients | Simple | Fixed Typically $6-9\,$V | None | $\frac{C_{blk}V_{th}}{I_{chg}}$ |
| DESAT with $R_e$ Fig. 11(a) | Tunable with $R_e$ | Good | Simple | Fixed Typically $6-9\,$V | $R_e$ | $R_e C_{blk} \ln \frac{V_{OUT} + I_{chg}R_e}{V_{OUT} + I_{chg}R_e - V_{th}}$ |
| DESAT with Type-2 IC Fig. 13 | Tunable with $R_1$, $R_2$, $R_3$ | Good | Moderate (increased passive components) | Programmable using $R_2$ and $R_3$ | $R_1$ | $R_{eq}C_{blk} \ln\left(1 - \frac{R_1 + R_2 + R_3}{R_3}\frac{V_{th}}{V_{OUTH}}\right)$ |

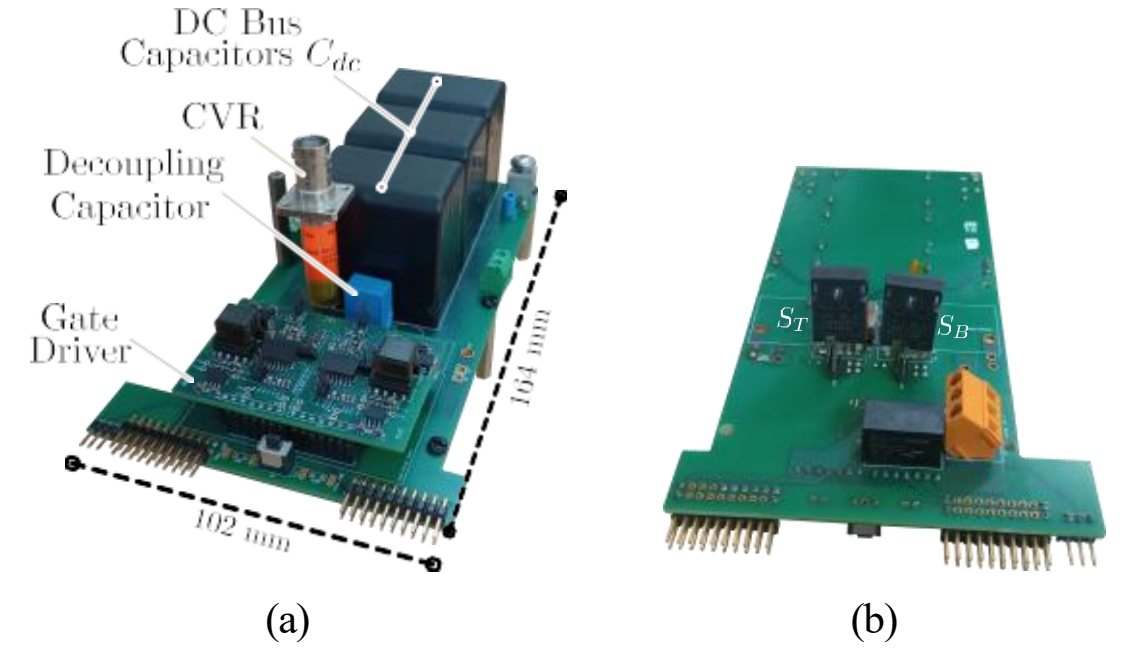

(a) (b)

Fig. 17. DPT setup for SiC MOSFET (a) top view, (b) bottom view.

Furthermore, Table III shows a comparison of various parameters between the three types of DESAT circuit configurations as shown in Fig. 5(a), Fig. 11(a) and Fig. 13.

## VIII. EXPERIMENTAL VERIFICATION

To verify the analysis experimentally, a double pulse test (DPT) circuit developed in [25] is utilized in this article, as shown Fig. 17. The experimental test conditions are set to $V_{dc} = 400$ V and load current of 25 A. To accurately capture the experimental results for SiC MOSFETs, high-bandwidth measurement equipment is employed. Specifically, $v_{ds}$ and $v_{C_{blk}}$ are measured using an optically isolated 500 MHz TIVP05 differential probe equipped with a 10X and 500X measuring tips. A Tektronix oscilloscope MSO44 with 1 GHz bandwidth is used to capture the results. $i_{ds}$ is measured using a T&M Research's SDN-414-025 current viewing resistor (CVR) with 1.2 GHz bandwidth. Two gate drivers are designed using ACPL337J and UCC21732 ICs to validate the two types of DESAT protection architectures shown in Fig. 11(a) and Fig. 13. The following subsections discuss the experimental performance and validation of both DESAT protection architectures in detail.

### *A. Experimental Results and Discussion on Type-1 DESAT Method*

A gate driver is designed using the ACPL337J IC which is also used in [23], which has the $V_{th}$ and $I_{chg}$ specifications that closely match the parameters evaluated in the quantitative analysis, as shown in Table II. A total gate resistance of $7.5\,\Omega$ with 15 V/ − 3 V gate voltages is selected to limit the gate driver output peak current to 2.5 A, as recommended in the datasheet. Further, an RS1MDF-13 diode ($D_d$) with a typical $C_j = 5$ pF is utilized. The voltage $v_{C_{blk}}$ is measured during normal turn-off of the DUT, as shown in Fig. 18 and 19, for two extreme $C_{blk}$ values such as 100 pF and 700 pF. The values of $R_d$ and $C_j$ are also varied to evaluate their impact on $v_{C_{blk}}$.

*1) During normal turn-off transients:* Fig. 18(a) shows the measured $v_{ds}$, $i_{ds}$ and $v_{C_{blk}}$ waveforms during turn-off for $R_d = 500\,\Omega$ and $C_{blk} = 100$ pF, where the measured $dv_{ds}/dt$ is approximately 40 V/ns. Under this condition, the peak transient voltage of $v_{C_{\text{blk}}}$ reaches 19 V, which far exceeds $V_{th}$ as well as the absolute maximum DESAT pin voltage rating of the IC. As shown in Fig. 18(b), increasing $R_d$ to 1 kΩ results only a 1 V reduction in the peak $v_{C_{\text{blk}}}$, which shows a weak impact of $R_d$. Further, to evaluate the effect of $C_j$, three $D_d$ diodes are connected in series to reduce the equivalent junction capacitance. As observed in Fig. 18(c), this decreases peak voltage to 15 V. However, the peak of $v_{C_{\text{blk}}}$ still remains well above the desired window and near the maximum DESAT pin-voltage rating.

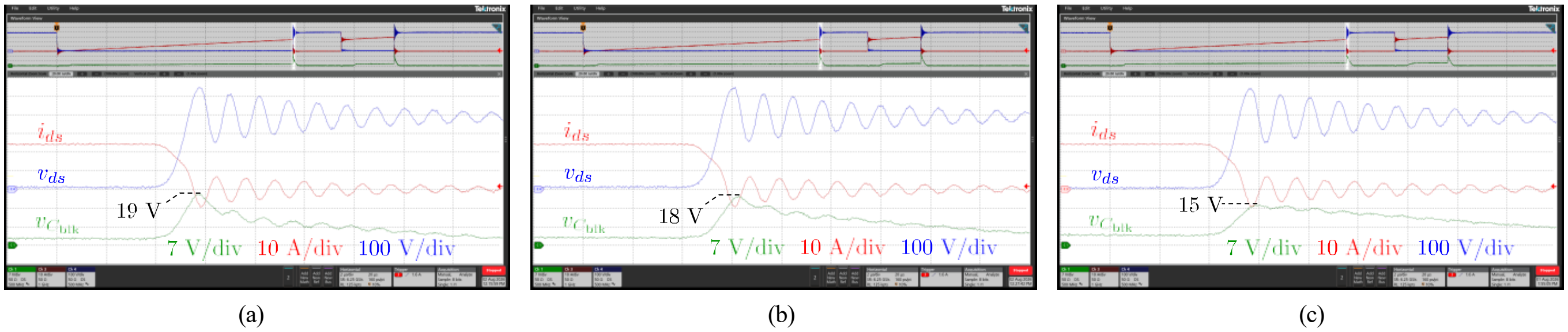

Fig. 18. Experimental results of $v_{C_{\text{blk}}}$ during normal turn-off transitions at $dv_{ds}/dt$ = 40 V/ns and $C_{\text{blk}}$ = 100 pF with (a) $R_d$ = 500 Ω with one $D_d$, (b) $R_d$ = 1 kΩ with one $D_d$, and (c) $R_d$ = 1 kΩ with three number of series connected $D_d$.

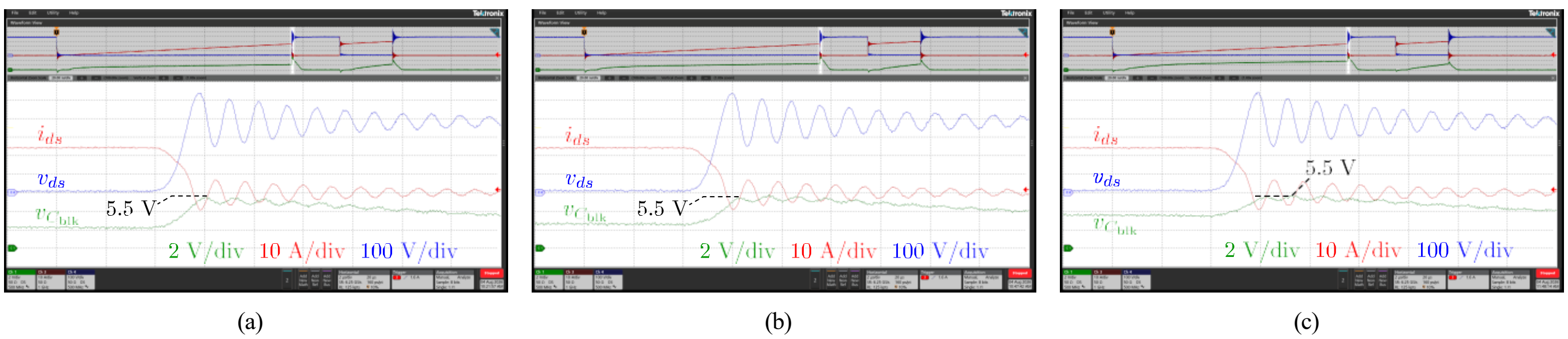

Fig. 19. Experimental results of $v_{C_{\text{blk}}}$ during normal turn-off transitions at $dv_{ds}/dt$ = 40 V/ns and $C_{\text{blk}}$ = 700 pF with (a) $R_d$ = 500 Ω with one $D_d$, (b) $R_d$ = 1 kΩ with one $D_d$, and (c) $R_d$ = 1 kΩ with three number of series connected $D_d$.

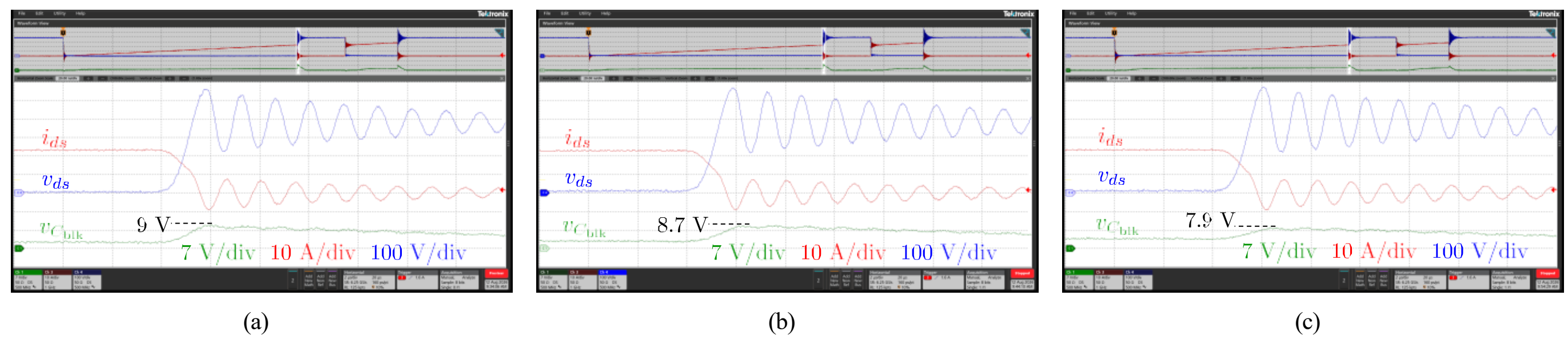

Fig. 20. Experimental results of $v_{C_{\text{blk}}}$ during normal turn-off transitions at $dv_{ds}/dt$ = 40 V/ns and $C_{\text{blk}}$ = 350 pF with (a) $R_d$ = 500 Ω with one $D_d$, (b) $R_d$ = 1 kΩ with one $D_d$, and (c) $R_d$ = 1 kΩ with three number of series connected $D_d$.

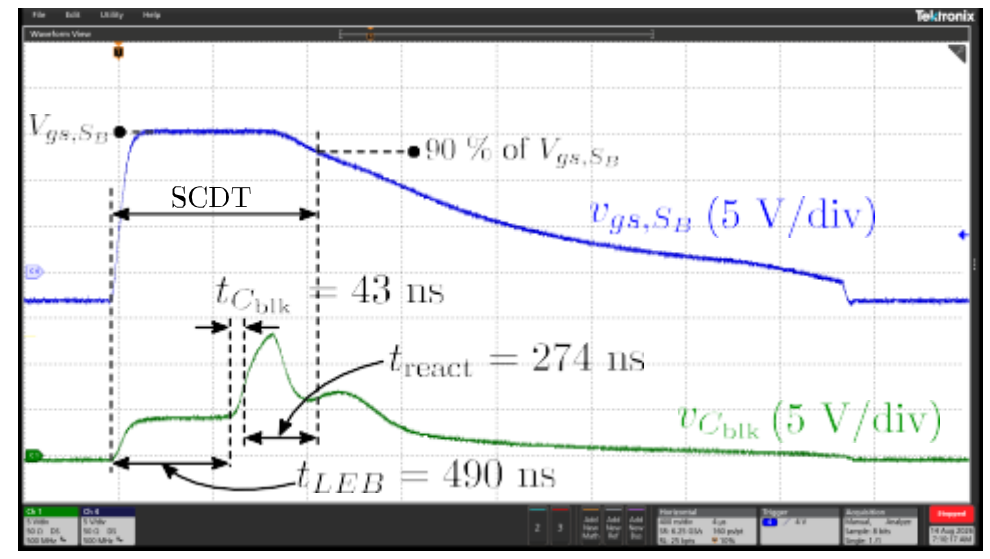

Fig. 21. Short-circuit emulation test result with ACPL337J for $C_{\text{blk}}$ = 350 pF and $R_e$ = 250 Ω.

Conversely, Fig. 19(a) shows the waveforms for $C_{\text{blk}}$ = 700 pF and $R_d$ = 500 Ω using single diode $D_d$. With this larger value of $C_{\text{blk}}$, the measured peak voltage of $v_{C_{\text{blk}}}$ is 5.5 V which remains safely below the $V_{th}$. Similarly, Fig. 19(b) and Fig. 19(c) show the results with $R_d$ = 1 kΩ and three series-connected diodes, respectively. It can be observed that the peak voltage remains unchanged with $R_d$ and $C_j$, which is also evident from the analysis shown in Fig. 9 for large $C_{\text{blk}}$ values.

However, since the main objective of this article is to select a value of $C_{\text{blk}}$ that attenuates the peak voltage $1-2$ V above $V_{th}$, a value of $C_{\text{blk}}$ = 350 pF is finally selected from the analytical plots shown in Fig. 9. To validate this selection, hardware tests were conducted at $C_{\text{blk}}$ = 350 pF under identical $R_d$ and diode configurations, as shown in Fig. 20.

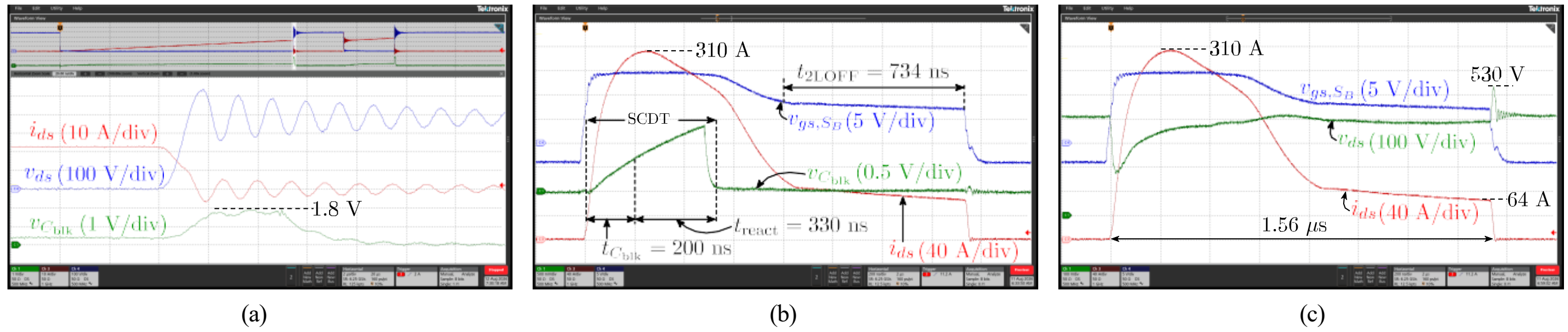

Fig. 22. Experimental results showing (a) $v_{C_{\rm blk}}$ during normal turn-off transition with $dv_{ds}/dt$ = 40 V/ns, $R_1$ = 600 Ω, $R_2$ = 7 kΩ, $R_3$ = 1.45 kΩ, and $C_{\rm blk}$ = 500 pF, (b) $v_{C_{\rm blk}}$, gate voltage ($v_{gs,S_B}$) of the DUT, and current through DUT ($i_{ds}$) during hard-switching fault, (c) voltage across the DUT ($v_{ds}$) and $i_{ds}$ during hard-switching fault.

TABLE IV
TABLE REPRESENTING THE COMPARISON BETWEEN THE EXISTING LITERATURE AND THE ADVANTAGE OF THE WORK CARRIED OUT IN THIS ARTICLE

| Literature | Type of DESAT Circuit | Value of $C_{\rm blk}$ | SCDT | Peak Transient Voltage of $v_{C_{\rm blk}}$ due to $dv_{ds}/dt$ during normal turn-off | Detection Threshold Voltage |
|---|---|---|---|---|---|
| [1] | Type-1 | 24 pF | 500 ns | Due to lower values of $C_{\rm blk}$, chosen only to satisfy SCDT requirement, peak transient voltage of $v_{C_{\rm blk}}$ induced by $dv_{ds}/dt$ may approach or exceed maximum allowable DESAT pin voltage rating. This is evident from the analysis presented in this article and experimental results shown in Fig. 18, which leads to IC damage or false fault detections [1], [2], and [13]. | 7 V |
| [2] | Type-1 with external resistors (no $I_{\rm chg}$) | - | 210 ns | | 5 V |
| [12] | Type-1 | 125 pF | ≈ 480 ns | | 2.64 V |
| [13], [14] | Type-1 with external resistors (no $I_{\rm chg}$) | 47 − 60 pF | 190 ns | | 3.6 V |
| [24] | Type-1 with external resistor | 200 pF | 400 − 600 ns | | 8 V |
| In this article | Type-1 with external resistor | 350 pF | 807 ns | The value of $C_{\rm blk}$ is selected to balance $dv_{ds}/dt$ noise and SCDT; peak transient $v_{C_{\rm blk}}$ safely remains within 1 − 2 V above $V_{\rm th}$ (Fig. 20 and 22(a)). | 7 V |
| | Type-2 | 500 pF | 530 ns | | 3.5 V |

With $R_d$ = 500 Ω and a single $D_d$, the peak $v_{C_{\rm blk}}$ voltage reaches 9 V, which falls around $V_{th}$. Similarly, increasing $R_d$ to 1 kΩ reduces the peak voltage to 8.7 V and employing three series diodes further reduces the peak voltage to 7.9 V, respectively.

*2) During HSF condition:* Furthermore, to meet the SCDT requirement, a value of $R_e$ = 250 Ω is selected from the plot shown in Fig. 12 for a 350 pF blanking capacitance. The corresponding short-circuit experimental result is shown in Fig. 21. It is observed that the measured SCDT is approximately 200 ns more than the target SCDT. Even though $C_{\rm blk}$ charges in 43 ns, the fixed inherent long $t_{LEB}$ of the IC dominates the total SCDT. Therefore, type-1 gate driver ICs with inherent longer delays are not suitable for SiC MOSFET applications. Further, the gate driver IC shuts down through a soft turn-off lasting approximately 2.2 $\mu$s followed by fault detection.

Although from the experimental results shown in Fig. 18-20, the trends of the peak $v_{C_{\rm blk}}$ voltage with respect to $R_d$ and $C_j$ follow the analytical predictions shown in Fig. 9, a deviation of 4 V to 5 V is observed in peak voltage of $v_{C_{\rm blk}}$ between the theoretical analysis and experimental measurements. This difference arises because of neglecting the (i) non-linear voltage dependence of $C_j$, (ii) reverse recovery current effect of $D_d$, and (iii) high-frequency parasitics of $R_d$.

### *B. Experimental Results and Discussion on Type-2 DESAT Method*

A gate driver is designed using the UCC21732 IC, utilizing the same gate resistance (7.5 Ω), gate bias voltages (15 V/ − 4 V), and DESAT diode ($D_d$) as the type-1 setup.

*1) During normal turn-off transient:* Fig. 22(a) shows the measured $v_{C_{\rm blk}}$ waveform during normal device turn-off for $R_1$ = 600 Ω, $R_2$ = 7 kΩ, $R_3$ = 1.45 kΩ, and $C_{\rm blk}$ = 500 pF, as designed in Section VII. Under a measured $dv_{ds}/dt$ of 40 V/ns, the peak $v_{C_{\rm blk}}$ transient reaches 1.8 V, which exceeds $V_{th}$ by approximately 1 V and lies in the desired window.

*2) During HSF condition:* Fig. 22(b) and 22(c) show the corresponding hard-switching short-circuit test result at $V_{dc}$ = 400 V. It can be observed that the total SCDT is 530 ns, which satisfies the target SCDT limit. Further, the gate driver turns off the DUT via a two-level turn-off after detecting the HSF. The two-level turn-off duration is measured as $t_{\rm 2LOFF}$ = 734 ns. Finally, the DUT is fully turned off within the time interval of 1.56 $\mu$s.

Therefore, $C_{\rm blk}$ = 500 pF effectively balances peak transient voltage attenuation according to 1 − 2 V above $V_{th}$ desired window and short-circuit protection speed for SiC MOSFETs. Further, the temperature-dependent short-circuit tests are performed at higher case temperatures, analyzed in detail, and reported in [16].

Furthermore, Table IV presents a comparison between the literature and the work reported in this article. It can be noticed that, with a type-2 DESAT circuit, even though a large $C_{\rm blk}$ value of 500 pF achieves the desired SCDT suitable for SiC MOSFET applications, simultaneously attenuating the peak transient voltage to the desired window.

## IX. Conclusion

This article presented a detailed design analysis of three DESAT protection circuit variants, followed by a quantitative analysis of $dv_{ds}/dt$ induced noise effects for each configuration. This analysis reveals that the conventional type-1 DESAT circuit topology fails to simultaneously achieve $dv_{ds}/dt$ noise immunity and a SCDT suitable for fast-switching SiC MOSFETs. To overcome this limitation, a type-1 DESAT circuit with external resistance, and a type-2 DESAT topology are shown to successfully maintain the peak transient voltage across $C_{\text{blk}}$ during $dv_{ds}/dt$ within a recommended window of $1 - 2$ V above the internal threshold voltage, while concurrently satisfying SCDT for SiC MOSFETs. Crucially, operating within this design window prevents possible false fault detection while ensuring that the peak transient voltage across $C_{\text{blk}}$ remains safely below the maximum voltage rating of the gate driver IC's DESAT pin. Further type-1 gate driver ICs with inherent shorter delays should be preferred.

Quantitatively, this optimal trade-off is achieved by selecting a DESAT diode with $C_j \leq 5$ pF, alongside $C_{\text{blk}} = 350$ pF with $R_e \leq 500\,\Omega$ for type-1 DESAT circuit with external resistance and $C_{\text{blk}} = 500$ pF with $R_1 = 600\,\Omega$ for type-2 circuit. Under these parameters, experimental measurements demonstrate that the transient peak voltage across $C_{\text{blk}}$ remains at 51 % − 58 % of the maximum DESAT pin rating for the type-1 gate driver IC. While it remains at 30 % for type-2 IC, providing a substantial overvoltage headroom by preserving the targeted SCDT. Finally, under hard-switching short-circuit fault conditions, the chosen values from the analysis yield a measured SCDT of 530 ns, which meets the design target and fully verifies the capability to reliably protect SiC MOSFETs against short-circuit faults.

## Acknowledgments

This work is supported by the Ministry of Earth Sciences, New Delhi, India, under the Grant No: MOES/PAMC/DOM/22-II/2022 (E-13674).